\documentclass[useAMS]{mn2e}
\usepackage{rotate}
\usepackage{rotating}
\usepackage{times}
\usepackage{verbatim}
\usepackage{graphicx}
\newif\ifAMStwofonts
\AMStwofontstrue

\def\xmm{{\it XMM-Newton }}

\title[Principal component analysis of blazars]
      {X-ray spectral variability of blazars using principal component analysis}

\author[D. Gallant et al.]
       {D. Gallant,$^1$ 
       L. C. Gallo,$^1$
       and M. L. Parker$^{2}$
       \\
$^{1}$ Department of Astronomy and Physics, Saint Mary's University, 923 Robie Street, Halifax, NS, B3H 3C3, Canada \\
$^{2}$ European Space Astron Ctr ESA ESAC, E-28691 Madrid, Spain \\
}
\date{Accepted. Received. }
\pagerange{\pageref{firstpage}--\pageref{lastpage}}
\pubyear{2017}
\begin{document}
\maketitle

\begin{abstract}
Principal Component Analysis (PCA) is applied to a variety of blazars to examine X-ray spectral variability. Data from nine different objects are analysed in two ways: long-term, which examines variability trends across years or decades, and short-term, which looks at variability within a single observation. The results are then compared to simulated spectra in order to identify the physical components that they correspond to. It is found that long-term variability for all objects is dominated by changes in a single power law component. The primary component is responsible for more than 84 per cent of the variability in every object, while the second component is responsible for at least 3 per cent. Small differences in the shapes of these components can be used to predict qualities such as the degree to which spectral parameters are varying relative to one another, and correlations between spectral hardness and flux. Short-term variability is less clear-cut, with no obvious physical analogue for some of the PCA results. We discuss the simulation process, and specifically remark on the consequences of the breakdown of the linearity assumption of PCA and how it manifests in the real data. We conclude that PCA is a useful tool for analysing variability, but only if its underlying assumptions and limitations are understood.

\end{abstract}


\section{Introduction}

Principal component analysis (PCA) is a model-independent technique that performs a change of basis on a data set, converting it into a series of orthogonal eigenvectors that best describe the variability within that data. These vectors are known as the principal components. The number of principal components is the minimum of $n -1$ and $v$, where $n$ is the number of observations in the original data set and $v$ is the number of variables in each observation. This means that most of principal components them will be unnecessary. The components are therefore ranked in order of their contribution to the total variability, which allows us to tell which ones are significant and which are not. The main benefit of this technique is that it can reduce redundancies within the data by expressing a potentially messy data set in terms of only a few fundamental trends in a model-independent way. Given a large enough data set, a series of X-ray spectra in our case, the physical parameters underlying the data can be reproduced with a high degree of accuracy. This can reveal processes going on deep within an active galactic nucleus (AGN) even when spectral fitting can not distinguish between two or more models, or provide a closer look at an object's variability independent of other factors (e.g. Kendall 1975). \\

PCA is used in many academic fields, and has recently seen significant application to X-ray astronomy. Francis $\&$ Willis (1999) provide an introduction to PCA of AGN, while Grupe et al. (1999) and Grupe (2004) demonstrate some early uses of PCA applied to ROSAT data. More recently, PCA has been applied to \xmm data to examine X-ray spectral variability in detail (for example, Vaughan $\&$ Fabian 2004, Miller et al. 2007, and Turner et al. 2007). In addition to AGN, PCA has been used to study variability in objects such as X-ray binaries (Malzac et al. 2006; Koljonen et al. 2013). Today, with almost two decades of high-quality \xmm data available, PCA can be used to reveal variability trends within AGN over long timescales, as in Parker et al. (2015) \\

Parker et al. (2015) applied principal component analysis to a wide range of AGN, mostly radio-quiet, looking at long-term variability with \xmm data. They found that most objects displayed variability in a power law continuum, but prominent variations in reflection components (in MCG-6-30-15, NGC 4051, 1H0707-495, NGC 3516, and Mrk 766) and partial covering absorption (in NGC 4395, NGC 1365, and NGC 4151) were also common. Their AGN displayed between three and five principal components, with evidence for many qualitatively different variability mechanisms. Some other sources can show long-term changes associated with emission from the distant torus (e.g. Gallo et al. 2015). \\

This work applies a similar analysis to a smaller sample of objects, specifically blazars, across both long and short timescales. Blazars were chosen due to their simple spectra and rapid variability, as well as the lack of PCA results for many well-known blazars. Blazar spectra are dominated by the effects of the jet, which follows a synchrotron self-Compton shape (Mastichiadis 1997), resulting in X-ray spectra that conform closely to a single power law model. By applying PCA to objects that are already known to be spectrally simple, we can better understand the intricacies of this technique and examine what drives blazar variability.\\

In Section 2, we describe our sample and data analysis. Section 3 presents the long-term, or multi-epoch, PCA results. These PCAs use all available observations of an object to describe variability over the span of years. Section 4 presents the short-term, or single-observation, results. These PCAs take a longer observation of an object and divide it into several spectra in order to observe variability over the span of hours. In Section 5, we discuss PCAs applied to simulated data in order to compare models to the real results. Lastly, Sections 6 and 7 summarize our results and present the conclusions. 

\section{Sample and Data Processing}
Objects were selected from among those in Costamante $\&$ Ghisellini (2002) with publicly available \xmm data (Jansen et al 2001), as well as the well-known object 3C 273. Observation dates ranged from May 2000 to May 2017. Only the highest signal-to-noise EPIC-pn instrument (Struder et al. 2001) was used. \\

Observations where the target was significantly off-axis, meaning the object was not near the centre of the field of view, were excluded. The data were collected in a variety of window modes and optical filters. Data in timing mode were not used, due to the uncertain calibration of this mode. For each observation, the observation data files (ODFs) were downloaded from the \xmm Science Archives and processed to create spectra using the Science Analysis System (SAS) version 15.0.0 \\

{\sc Epchain} was used to generate event lists from the ODFs, and the spectra were made using a source region with a radius of 35$\arcsec$. Background subtraction was performed using a background region of radius 50$\arcsec$ located near the source. \\

Each observation was checked for pileup, and some showed significant amounts of it. This was corrected for by extracting the source spectrum from an annulus with the same outer radius, and an inner radius of 8$\arcsec$, which excludes the most highly piled-up light from the centre of the object. To ensure that this corrective technique did not influence the results, PCAs of piled-up objects were performed both with and without the piled-up observations. Other than showing more noise due to the lower sample size, this caused no major difference in the shapes or significance of the principal components. \\

Some observations displayed high levels of background flaring at certain times. These observations were filtered through a good time interval (GTI) that excluded the times when the flaring occurred. Response matrices and ancillary response files were created using {\sc rmfgen} and {\sc arfgen}, respectively.\\

Figure 1 presents representative spectra for each object, unfolded against a power law with $\Gamma$ $=$ 0, where $\Gamma$ is the photon index of the power law. The complete list of observations is shown in Table 1.

\begin{table*}
\caption{Complete list of observations.}
\begin{center}
\begin{tabular}{ccccccccc}\hline
Object & Obs ID & Revolution & Start Time & Duration & GTI & 0.3-10 keV Count Rate $(s^{-1})$ & Pileup Correction? & Window Mode  \\ \hline \hline
3C 273 & 0112770101 & 370 & 2001-12-16 15:35:23  & 6399 & 3507 & 64.3 & N & Small \\
 & 0112770201 & 373 & 2001-12-22 00: 19:58 & 6399 & 3471 & 62.23 & N & Small \\
 & 0112770501 & 655 & 2003-07-08 10:33:51 & 8553 & 5631 & 62.67 & N & Small \\ 
 & 0112770601 & 472 & 2002-07-07 14:25:05 & 5996 & 3504 & 47.91 & N & Small\\
 & 0112770701 & 563 & 2003-01-05 17:24:04 & 5630 & 3503 & 58.32 & N & Small \\ 
 & 0112770801 & 554 & 2002-12-17 22:24:56 & 5624 & 3503 & 69.37 & N & Small \\
 & 0112771001 & 645 & 2003-06-18 01:07:13 & 5950 & 3861 & 70.72 & N & Small \\
 & 0112771101 & 735 & 2003-12-14 19:23:21 & 12849 & 5928 & 47.87 & N & Small \\
 & 0126700301 & 94 & 2000-06-13 23:39:53 & 73556 &  45260 & 42.02 & N & Small \\ 
 & 0126700601 & 95 & 2000-06-15 12:58:18 & 31032 & 20820 & 40.46 & N & Small  \\
 & 0126700701 & 95 & 2000-06-15 23:32:02 & 36346 & 21030 & 39.29 & N & Small \\
 & 0126700801 & 96 & 2000-06-17 23:24:14 & 73561 & 42510 & 45.52 & N & Small \\
 & 0136550101 & 277 & 2001-06-13 07:14:26 & 89765 & 62000 & 53.65 & N & Small\\
 & 0136550501 & 563 & 2003-01-05 14:17:24 & 8951 & 5965 & 66.58 & N & Small \\
 & 0136550801 & 835 & 2004-06-30 13:02:25 & 62913 & 13910 & 40.40 & N & Small \\
 & 0136551001 & 1023 & 2005-07-10 13:51:19 & 28111 & 19330 & 44.24 & N & Small \\
 & 0159960101 & 655 & 2003-07-07 17:40:27 & 58557 & 40600 & 63.48 & N & Small \\
 & 0414190101 & 1299 & 2007-01-12 07:13:55 & 78566 & 53710 & 49.47 & N & Small \\
 & 0414190301 & 1381 & 2007-06-25 05:08:14 & 32511 & 22440 & 40.83 & N & Small \\
 & 0414190401 & 1465 & 2007-12-08 20:11:25 & 35875 & 24820 & 81.17 & N & Small \\
 & 0414190501 & 1649 & 2008-12-09 20:12:31 & 41015 & 28420 & 57.01 & N & Small \\
 & 0414190601 & 1837 & 2009-12-20 03:42:44 & 31912 & 22030 & 62.46 & N & Small \\
 & 0414190701 & 2015 & 2010-12-10 01:37:45 & 36414 & 25210 & 46.88 & N & Small\\
 & 0414190801 & 2199 & 2011-12-12 17:44:21 & 43915 & 30380 & 42.22 & N & Small \\
 & 0414191001 & 2308 & 2012-07-16 11:59:23 & 38918 & 17760 & 36.70 & N & Small \\
 & 0414191101 & 2856 & 2015-07-13 21:03:55 & 72400 & 49680 & 31.73 & N & Small \\
 & 0414191201 & 3031 & 2016-06-26 20:22:08 & 67200 & 46030 & 55.51 & N & Small \\ \hline
3C 279 & 0651610101 & 2035 & 2011-01-18 16:49:52 & 126346 & 86960 & 49.48 & N & Small \\ \hline
H1426+428 & 0111850201 & 278 & 2001-06-16 00:49:21 & 68574 & 45770 & 16.83 & N & Small \\ 
 & 0165770101 & 852 & 2004-08-04 00:59:26 & 67866 & 45860 & 20.13 & N & Small \\
 & 0165770201 & 853 & 2004-08-06 00:32:43 & 68920 & 47980 & 20.06 & N & Small \\
 & 0212090201 & 939 & 2005-01-24 14:44:40 & 30417 & 20960 & 25.13 & N & Small \\
 & 0310190101 & 1012 & 2005-06-19 07:39:40 & 47034 & 32680 & 36.97 & N & Small \\
 & 0310190201 & 1015 & 2005-06-25 06:03:28 & 49505 & 31140 & 28.79 & N & Small \\
 & 0310190501 & 1035 & 2005-08-04 04:52:10 & 47542 & 32410 & 28.47 & N & Small\\ \hline
 Mrk 421 & 0099280101 & 84 & 2000-05-25 03:17:11 & 66497 & 21160 & 216.8 & N & Small \\ 
 & 0099280201 & 165 & 2000-11-01 23:47:51 & 40115 & 24240 & 112.5 & Y & Small \\
 & 0099280301 & 171 & 2000-11-13 22:00:29 & 49811 & 25640 & 279.7 & N & Small\\
 & 0136540101 & 259 & 2001-05-08 09:09:35 & 39007 & 25730 & 144.5 & Y & Small \\
 & 0136540301 & 532 & 2002-11-04 00:44:59 & 23913 & 13830 & 23.40 & N & Full-frame\\
 & 0136540401 & 532 & 2002-11-04 07:41:43 & 23917 & 14180 & 43.51 & Y & Full-frame\\
 & 0136540701 & 537 & 2002-11-14 00:07:35 & 71520 & 37970 & 97.45 & Y & Large \\
 & 0153950601 & 440 & 2002-05-04 16:09:17 & 39727 & 34330 & 25.75 & Y & Large\\
 & 0153950701 & 440 & 2002-05-04 03:51:30 & 19982 & 15940 & 16.55 & Y & Large\\
 & 0158970101 & 637 & 2003-06-01 11:33:26 & 47538 & 24920 & 103.3 & Y & Small \\
 & 0162960101 & 733 & 2003-12-10 21:23:14 & 50755 & 16470 & 119.8 & Y & Small \\
 & 0411081301 & 1358 & 2007-05-10 03:37:41 & 18913 & 13960 & 37.50 & Y & Full-frame\\
 & 0411083201 & 1820 & 2009-11-16 17:37:59 & 58070 & 7526 & 112.3 & Y & Large \\
 & 0560980101 & 1640 & 2008-11-22 14:07:29 & 71318 & 8479 & 51.67 & Y & Large \\
 & 0560983301 & 1732 & 2009-05-25 03:37:32 & 64173 & 8468 & 63.64 & Y & Large \\
 & 0656380101 & 1904 & 2010-05-03 07:19:29 & 51169 & 6619 & 91.76 & Y & Large \\
 & 0656380801 & 2001 & 2010-11-12 20:51:05 & 42669 & 7628 & 66.98 & Y & Large \\
 & 0658800101 & 2094 & 2011-05-19 10:02:48 & 35074 & 8941 & 38.94 & Y & Large \\
 & 0658801301 & 2837 & 2015-06-05 23:48:35 & 29000 & 19270 & 105.6 & Y & Small \\
 & 0658801801 & 2915 & 2015-11-08 13:42:37 & 33600 & 21200 & 81.51 & Y & Small \\
 & 0658802301 & 3005 & 2016-05-06 03:38:20 & 29400 & 19540 & 72.91 & Y & Small \\
 & 0791780101 & 3096 & 2016-11-03 13:15:45 & 17500 & 11210 & 61.34 & N & Small \\
 & 0791780601 & 3187 & 2017-05-04 04:01:33 & 12500 & 7708 & 153.7 & N & Small \\ \hline
\end{tabular}
\end{center}
\end{table*}

\begin{table*}
\contcaption{}
\begin{center}
\begin{tabular}{ccccccccc}\hline
Object & Obs ID & Revolution & Start Time & Duration & GTI & Count Rate $(s^{-1})$ & Pileup Correction? & Window Mode  \\ \hline \hline
Mrk 501 & 0113060401 & 475 & 2002-07-14 17:02:39 & 15769 & 2945 & 0.1304 & N & Small \\
 & 0652570101 & 1969 & 2010-09-08 23:50:27 & 44912 & 31160 & 26.47 & N & Small \\
 & 0652570201 & 1970 & 2010-09-10 23:42:24 & 44919 & 31160 & 27.39 & N & Small \\
 & 0652570301 & 2047 & 2011-02-11 14:43:25 & 40914 & 28350 & 28.86 & N & Small \\
 & 0652570401 & 2049 & 2011-02-15 14:18:29 & 40715 & 28220 & 37.72 & N & Small \\ \hline
OJ 287 & 0300480201 & 978 & 2005-04-12 13:13:21 & 38913 & 9918 & 1.388 & N & Large \\
 & 0300480301 & 1081 & 2005-11-03 21:16:31 & 48059 & 28800 & 1.073 & N & Large\\
 & 0401060201 & 1271 & 2006-11-17 00:33:10 & 47211 & 41360 & 0.8658 & N & Large\\
 & 0502630201 & 1533 & 2008-04-22 17:13:34 & 55815 & 48100 & 0.8675 & N & Large\\
 & 0679380701 & 2170 & 2011-10-15 08:18:19 & 23917 & 20150 & 2.938 & N & Large \\
 & 0761500201 & 2822 & 2015-05-07 05:23:25 & 129200 & 94890 & 1.871 & N & Large \\ \hline
PG 1553+113 & 0656990101 & 1952 & 2010-08-06 12:38:17 & 21914 & 15050 & 15.19 & N & Small \\
 & 0727780101 & 2495 & 2013-07-24 14:57:49 & 34500 & 23120 & 28.98 & N & Small \\
 & 0727780201 & 2680 & 2014-07-28 04:00:06 & 36300 & 24380 & 17.25 & N & Small \\
 & 0727780301 & 2882 & 2015-09-04 18:23:24 & 29999 & 19960 & 10.24 & N & Small \\
 & 0727780401 & 3057 & 2016-08-17 21:56:06 & 30000 & 19960 & 12.33 & N & Small \\
 & 0761100101 & 2864 & 2015-07-29 19:57:33 & 138400 & 119700 & 5.569 & Y & Full-frame \\
 & 0761100201 & 2866 & 2015-08-02 19:40:00 & 138900 & 119000 & 4.575 & Y & Full-frame \\
 & 0761100301 & 2867 & 2015-08-04 19:32:00 & 138900 & 19960 & 10.24 & N & Small \\
 & 0761100401 & 2869 & 2015-08-08 19:12:07 & 138900 & 117700 & 4.348 & Y & Full-frame \\
 & 0761100701 & 2873 & 2015-08-30 18:52:06 & 90000 & 62010 & 8.622 & N & Small \\
 & 0761101001 & 2880 & 2015-08-30 17:52:29 & 139000 & 117200 & 5.798 & Y & Full-frame \\  \hline
PKS 2155-304 & 0080940101 & 174 & 2000-11-19 18:38:20 & 60511 & 40190 & 16.75 & Y & Small \\
 & 0080940301 & 174 & 2000-11-20 12:53:01 & 61411 & 40810 & 58.16 & N & Small \\
 & 0124930201 & 87 & 2000-05-31 00:30:51 & 72558 & 41580 & 77.13 & N & Small \\ 
 & 0124930301 & 362 & 2001-11-30 02:36:09 & 92617 & 31260 & 79.04 & Y & Small \\
 & 0124930501 & 450 & 2002-05-24 09:31:02 & 104868 & 22300 & 55.68 & N & Small \\
 & 0124930601 & 545 & 2002-11-29 23:27:28 & 114675 & 39790 & 29.71 & N & Small \\
 & 0158960101 & 724 & 2003-11-23 00:46:22 & 27159 & 18670 & 27.55 & N & Small \\
 & 0158960901 & 908 & 2004-11-22 21:35:30 & 28919 & 19960 & 30.90 & N & Small \\
 & 0158961001 & 908 & 2004-11-23 19:45:55 & 40419 & 27960 & 40.12 & N & Small \\
 & 0158961101 & 993 & 2005-05-12 12:51:06 & 28910 & 19250 & 69.63 & N & Small \\
 & 0158961301 & 1095 & 2005-11-30 20:34:03 & 60415 & 41900 & 76.16 & N & Small \\
 & 0411780101 & 1266 & 2006-11-07 00:22:47 & 101012 & 20870 & 42.28 & N & Small \\
 & 0411780201 & 1349 & 2007-04-22 04:07:23 & 67911 & 43360 & 74.47 & N & Small \\
 & 0411780301 & 1543 & 2008-05-12 15:02:34 & 61216 & 42600 & 89.11 & N & Small \\
 & 0411780401 & 1734 & 2009-05-28 08:08:42 & 64820 & 45100 & 62.32 & N & Small \\
 & 0411780501 & 1902 & 2010-04-29 20:26:00 & 74298 & 47730 & 31.87 & N & Small \\
 & 0411780601 & 2084 & 2011-04-26 13:50:40 & 63818 & 44400 & 49.04 & N & Small \\
 & 0411780701 & 2268 & 2012-04-28 00:48:26 & 68735 & 38660 & 12.56 & N & Small \\
 & 0411782101 & 2449 & 2013-04-23 22:31:38 & 76015 & 48830 & 27.62 & N & Small \\
 & 0727770901 & 2633 & 2014-04-25 03:14:56 & 65000 & 44500 & 29.68 & N & Small \\ \hline
S5 0716+714 & 0502271401 & 1427 & 2007-09-24 16:23:32 & 73917 & 50120 & 4.269 & N & Small  \\ \hline   
\end{tabular}
\end{center}
\end{table*}

\begin{figure*}
\begin{minipage}[]{0.30\hsize}
\scalebox{0.40}{\includegraphics[angle=0]{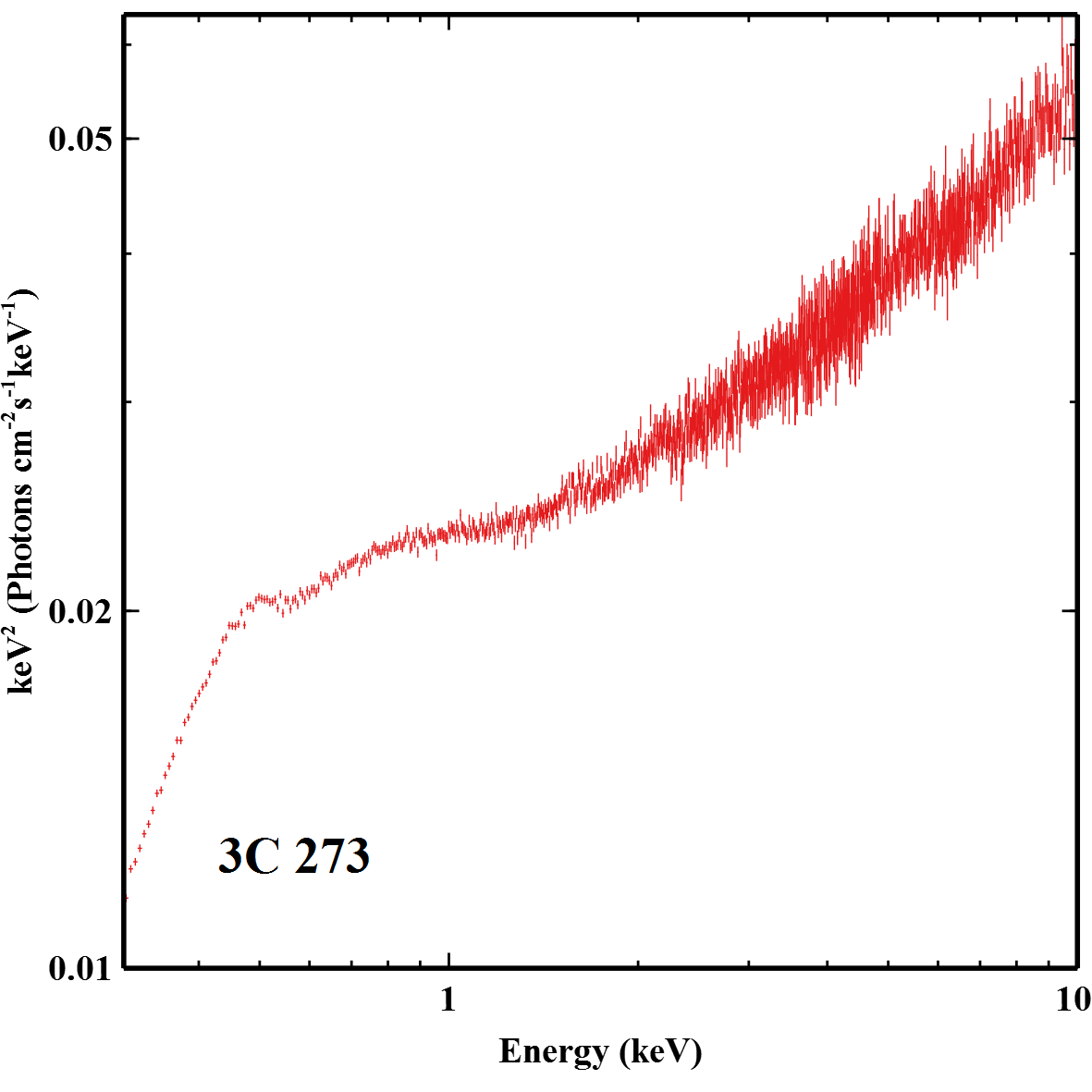}}
\end{minipage}
\hfill
\begin{minipage}[]{0.30\hsize}
\scalebox{0.40}{\includegraphics[angle=0]{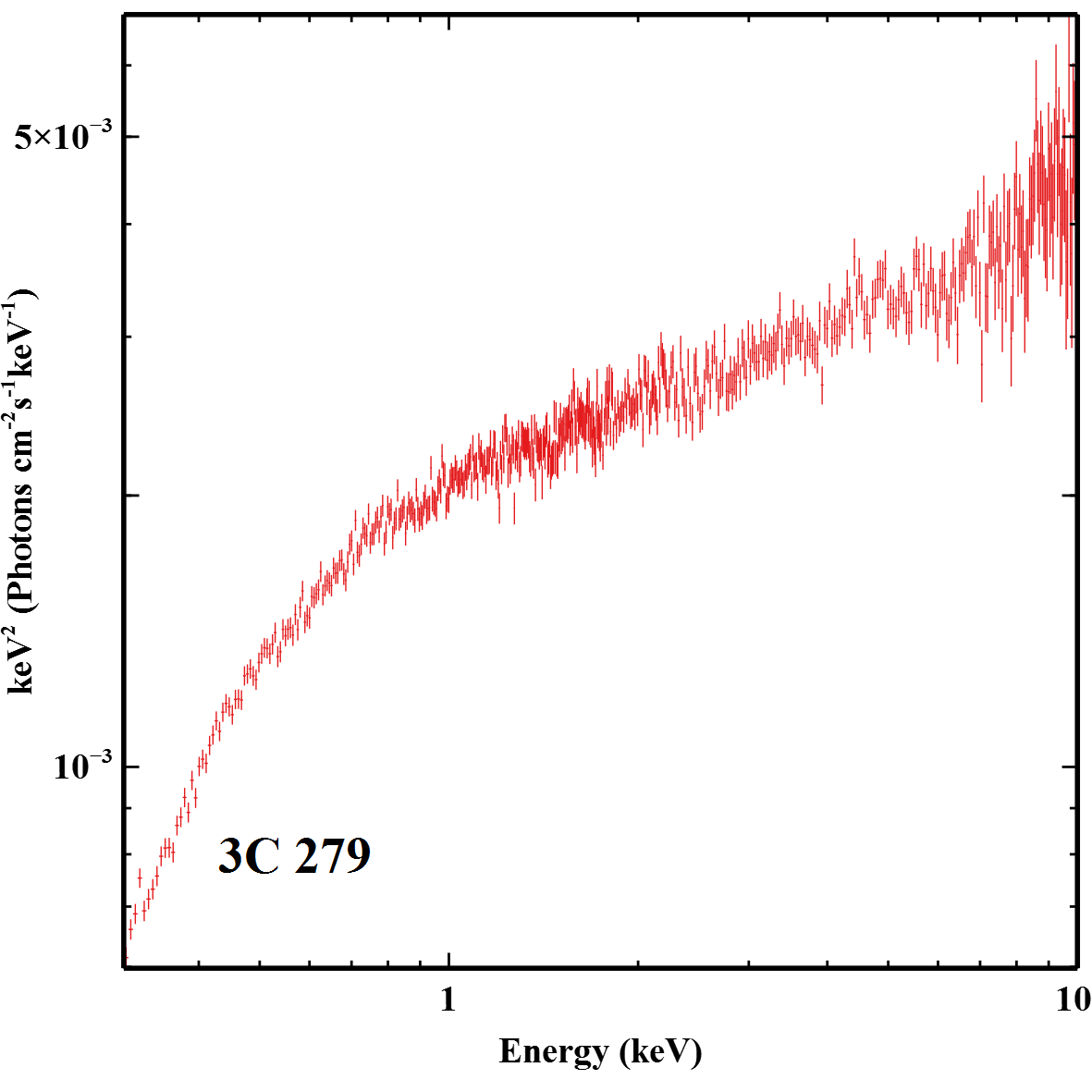}}
\end{minipage}
\hfill
\begin{minipage}[]{0.30\hsize}
\scalebox{0.40}{\includegraphics[angle=0]{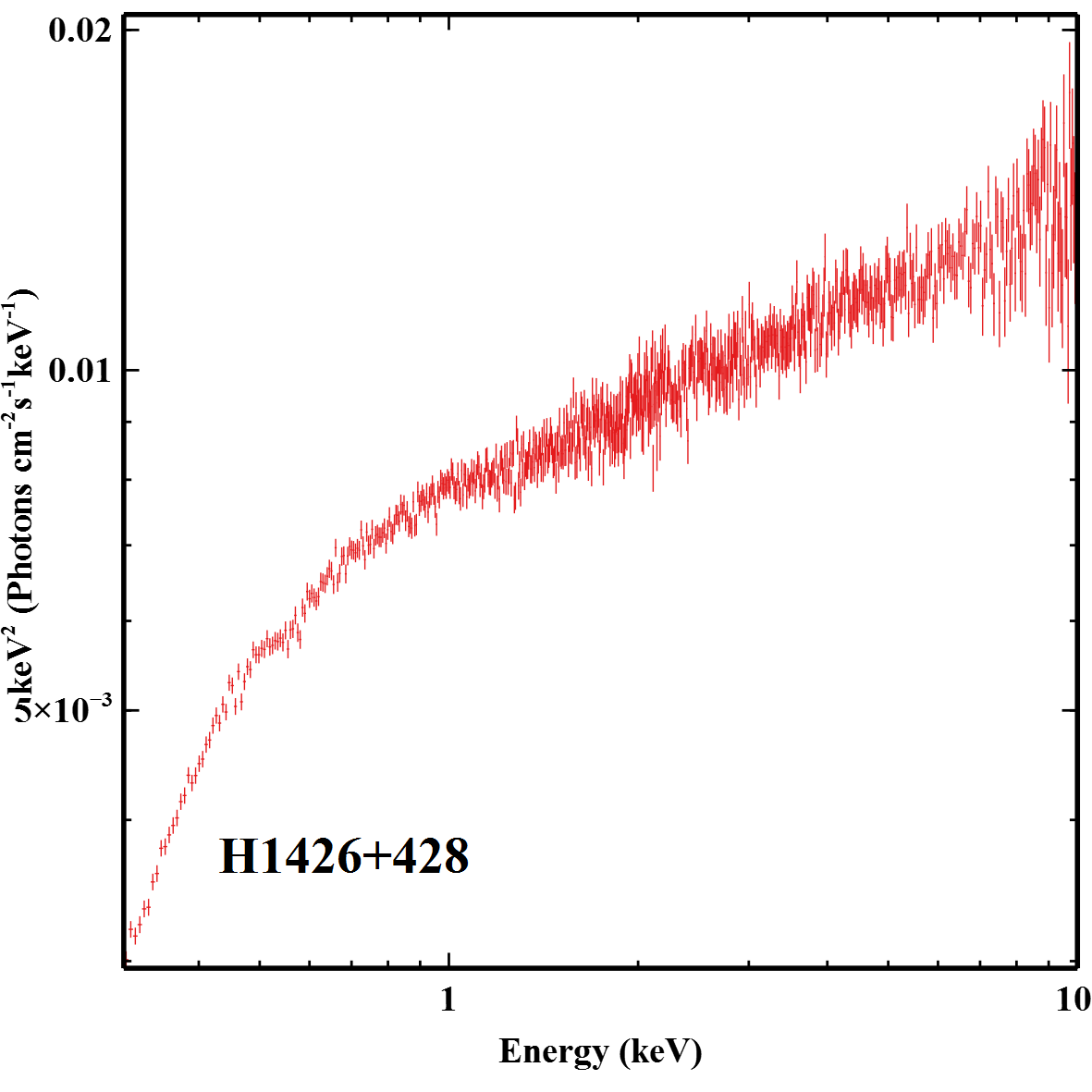}}
\end{minipage}
\hfill
\begin{minipage}[]{0.30\hsize}
\scalebox{0.40}{\includegraphics[angle=0]{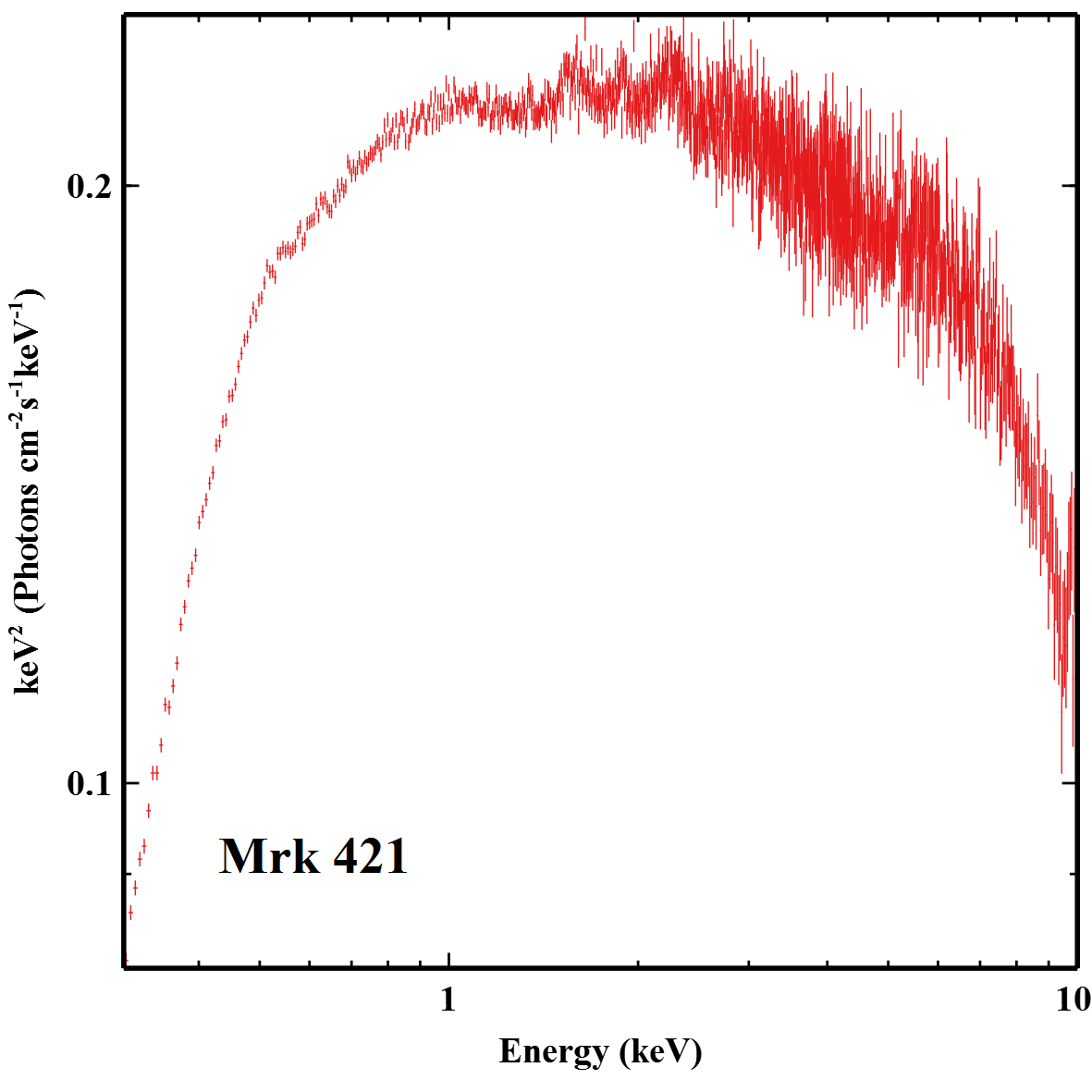}}
\end{minipage}
\hfill
\begin{minipage}[]{0.30\hsize}
\scalebox{0.40}{\includegraphics[angle=0]{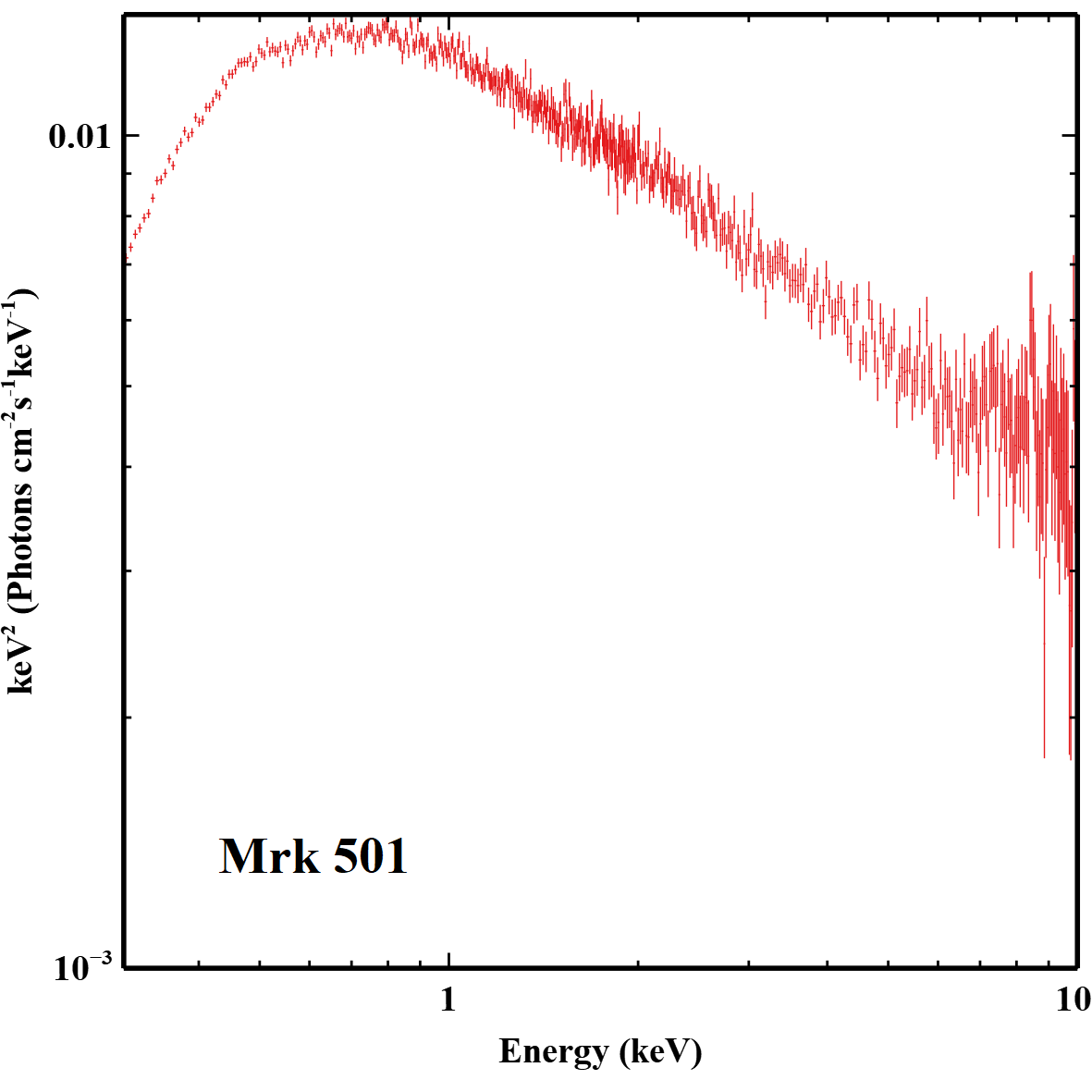}}
\end{minipage}
\hfill
\begin{minipage}[]{0.30\hsize}
\scalebox{0.40}{\includegraphics[angle=0]{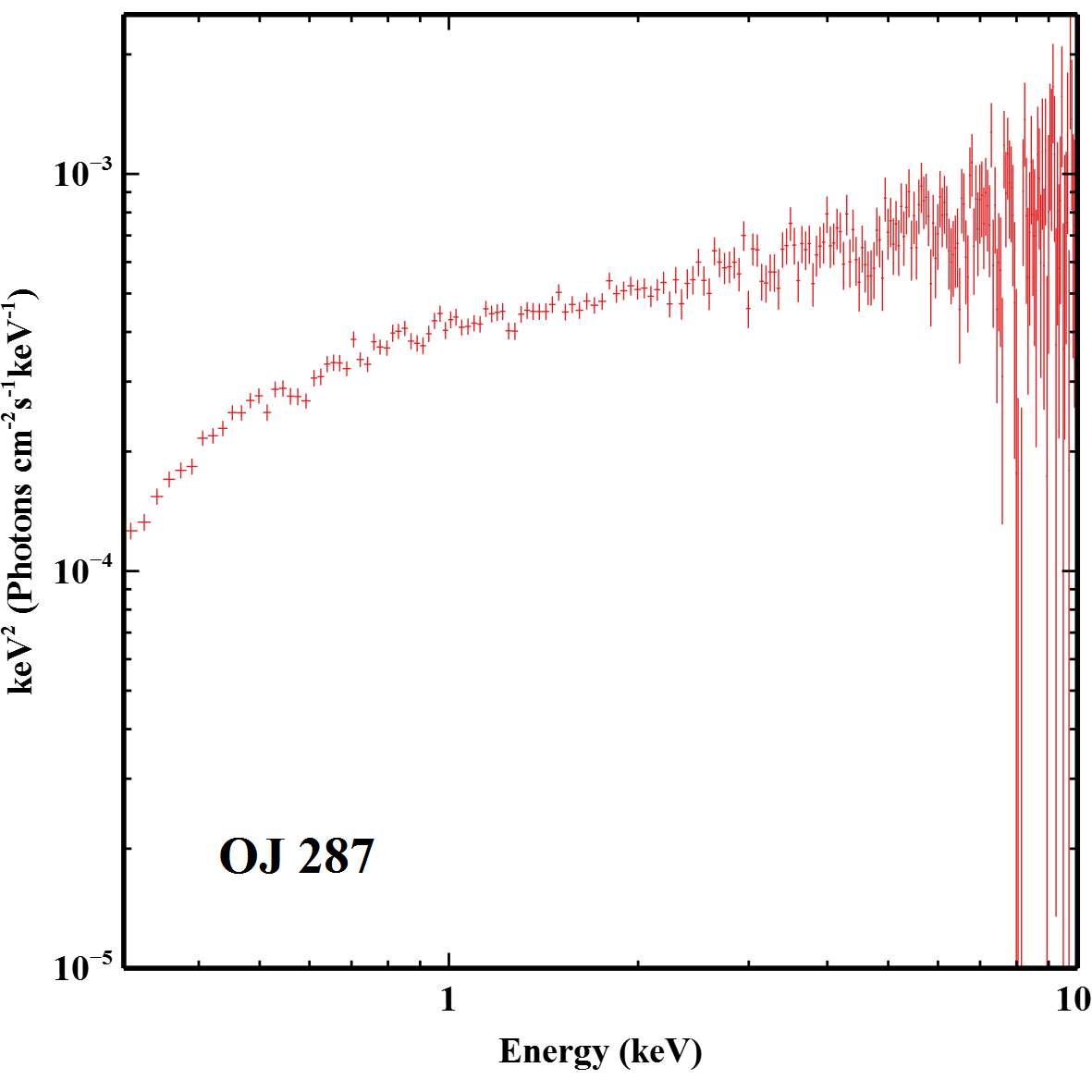}}
\end{minipage}
\hfill
\begin{minipage}[]{0.30\hsize}
\scalebox{0.40}{\includegraphics[angle=0]{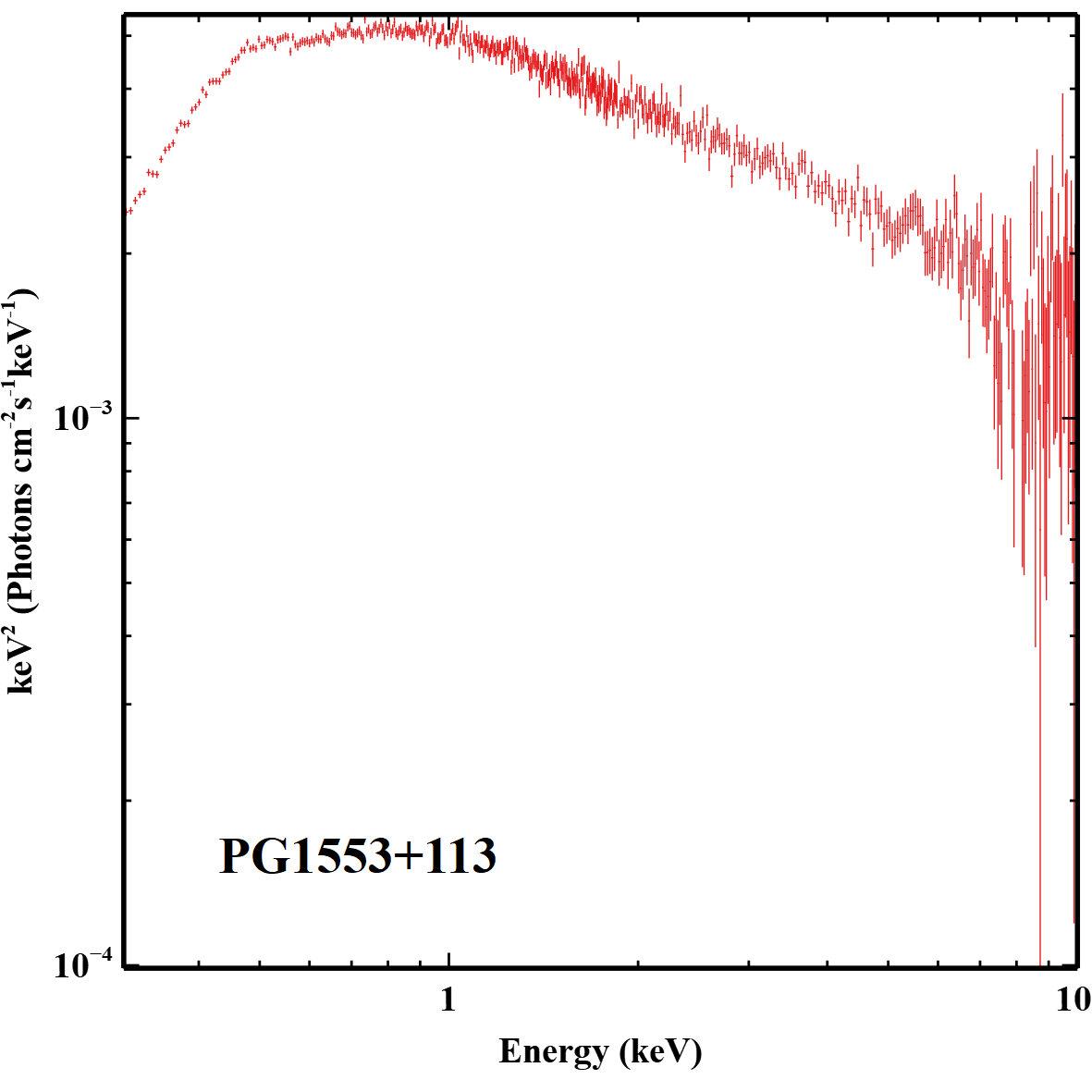}}
\end{minipage}
\hfill
\begin{minipage}[]{0.30\hsize}
\scalebox{0.40}{\includegraphics[angle=0]{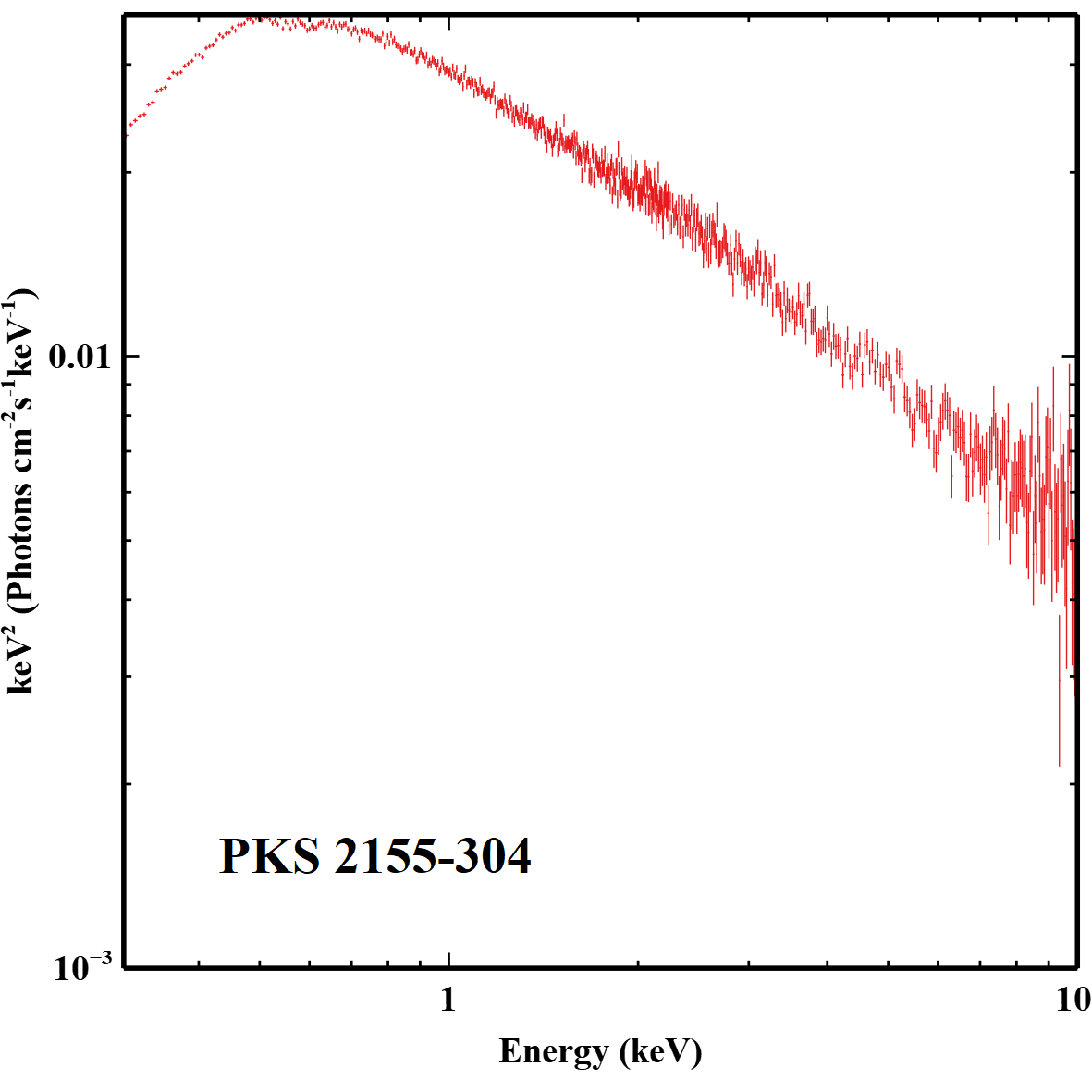}}
\end{minipage}
\hfill
\begin{minipage}[]{0.30\hsize}
\scalebox{0.40}{\includegraphics[angle=0]{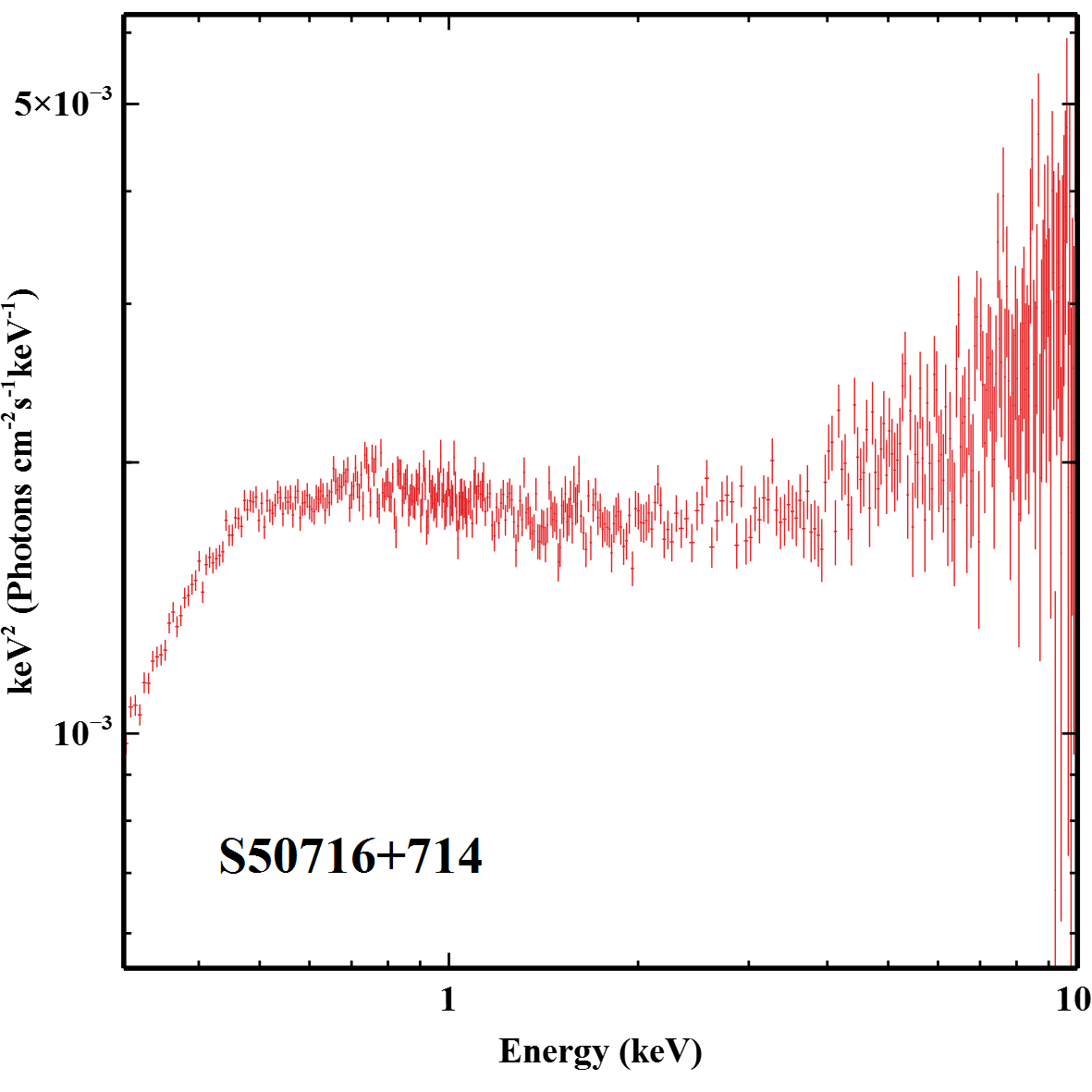}}
\end{minipage}
\hfill
\caption{Representative spectra for each object, unfolded against a power law with $\Gamma$ = 0.}
\end{figure*}

\section{Long-term (multi-epoch) variability}
This section presents the results of PCA performed on a single object across many different observations. For objects with more than four separate observations (of any duration), this PCA was calculated in order to examine long-term variability. Seven objects from our sample met this criteria: 3C 273, H1426+428, Mrk 421, Mrk 501, OJ 287, PG1553+113, and PKS 2155-304. \\

These observations span a minimum of four years (for H1426+428) to a maximum of seventeen years (for Mrk 421). The observations were not evenly spread out in time, with some sources being observed much more frequently than others. Each observation for a given object corresponded to a single spectrum used in the PCA. This is a departure from the method of Parker et al (2015), which divided each observation into 10ks parts. Their method captures both elements of long-term and short-term variability within the same PCA, whereas this work examines them separately by leaving the observations whole for the long-term analysis, and splitting them into segments for the short-term analysis in Section 4. 3C 273 was the most sampled object, with 27 observations over fifteen years. Figure 2 shows the spectral variability of this object by comparing all 27 observations to an average power law fit. \\

\begin{figure}
\begin{center}
\begin{minipage}{0.99\linewidth}
\scalebox{0.50}{\includegraphics[angle=0]{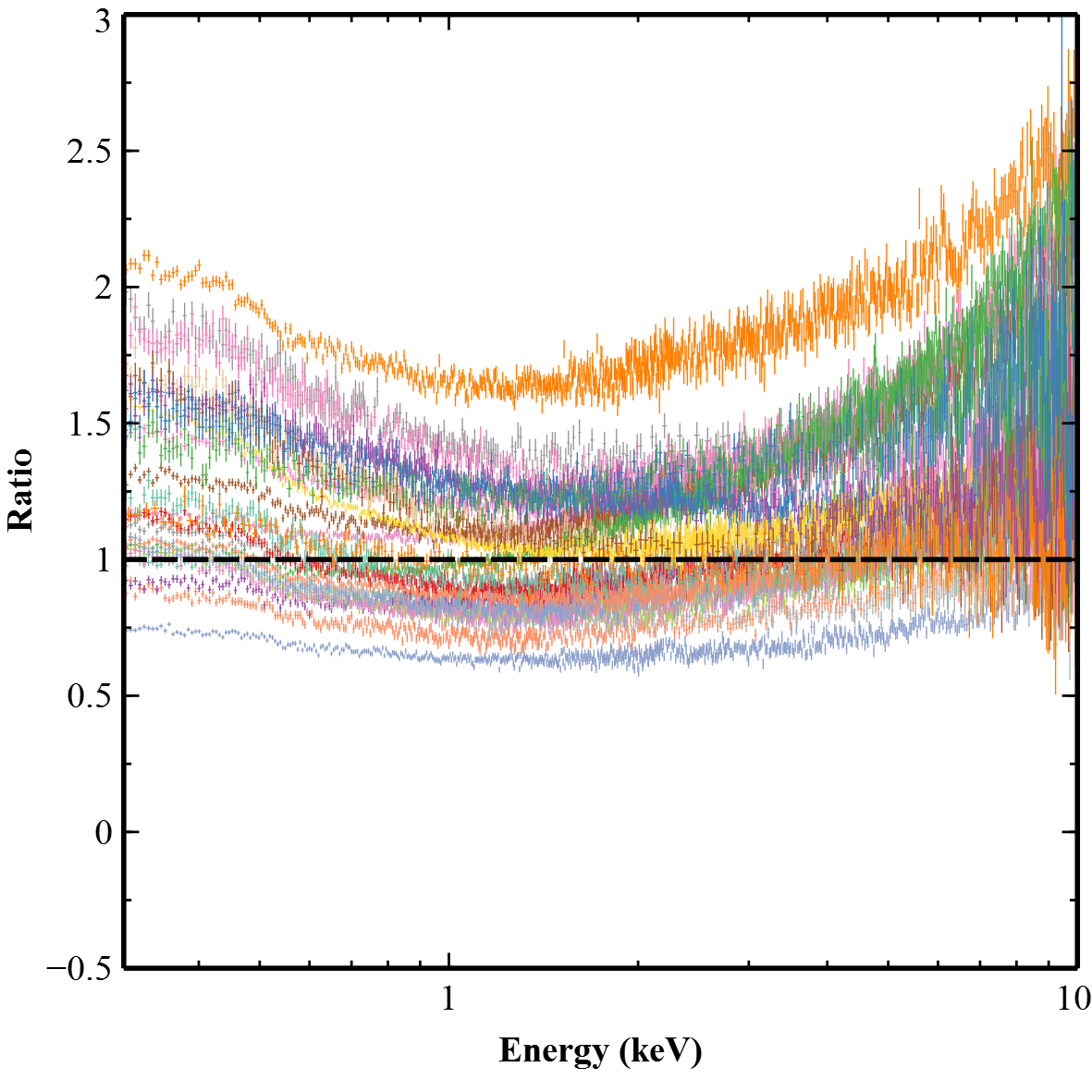}}
\end{minipage}  \hfill
\end{center}
\caption{Ratio of 27 spectra of 3C 273 spanning from 2001 to 2016 to an average power law fit. The fit was created by fitting a single power law model to every spectrum at once. This provides a visual example of long-term variability. This source is highly variable, both in the shape of the spectrum and the total flux. This is typical of blazars.}
\end{figure}

\begin{figure*}
\begin{minipage}[]{0.30\hsize}
\scalebox{0.40}{\includegraphics[angle=0]{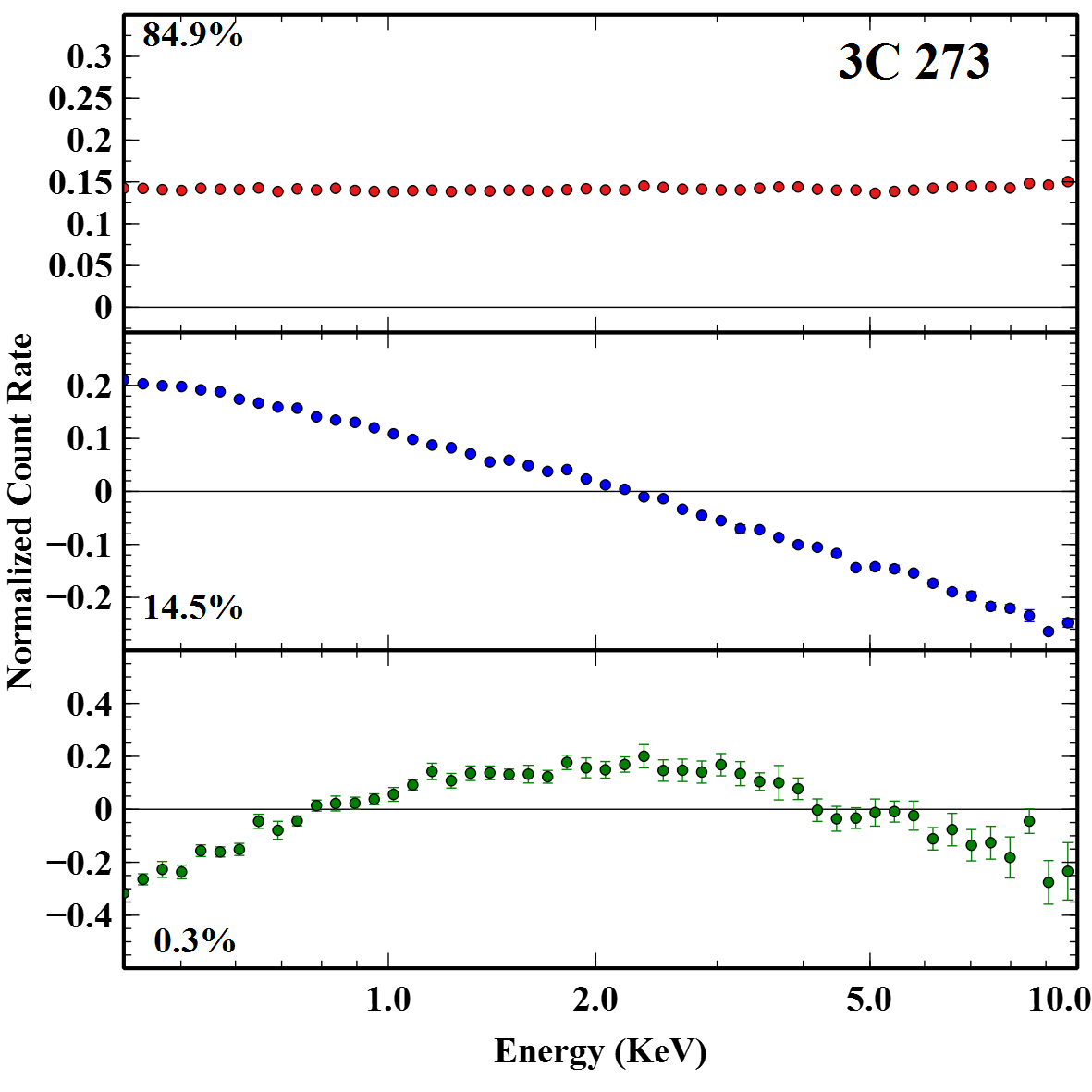}}
\end{minipage}
\hfill
\begin{minipage}[]{0.30\hsize}
\scalebox{0.40}{\includegraphics[angle=0]{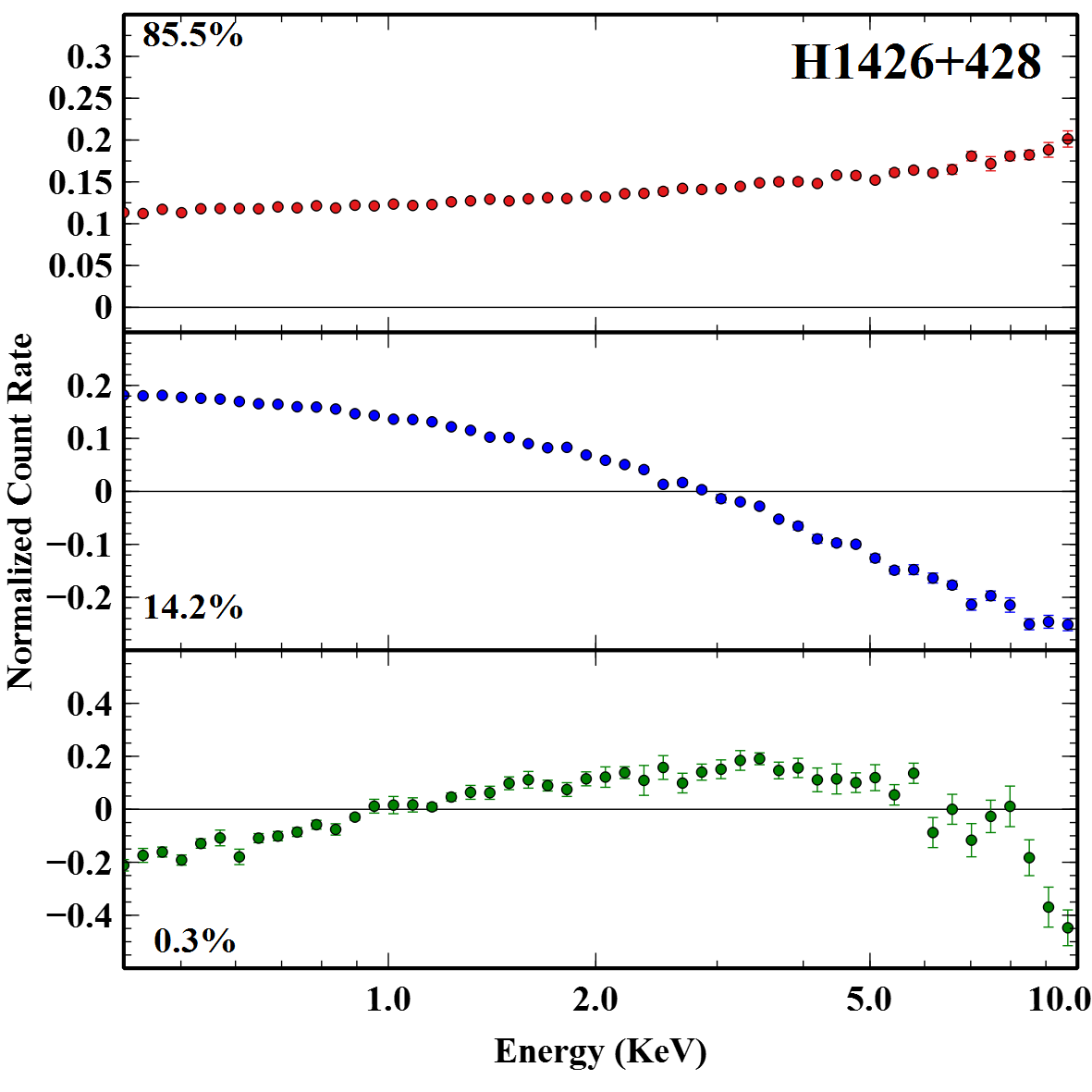}}
\end{minipage}
\hfill
\begin{minipage}[]{0.30\hsize}
\scalebox{0.40}{\includegraphics[angle=0]{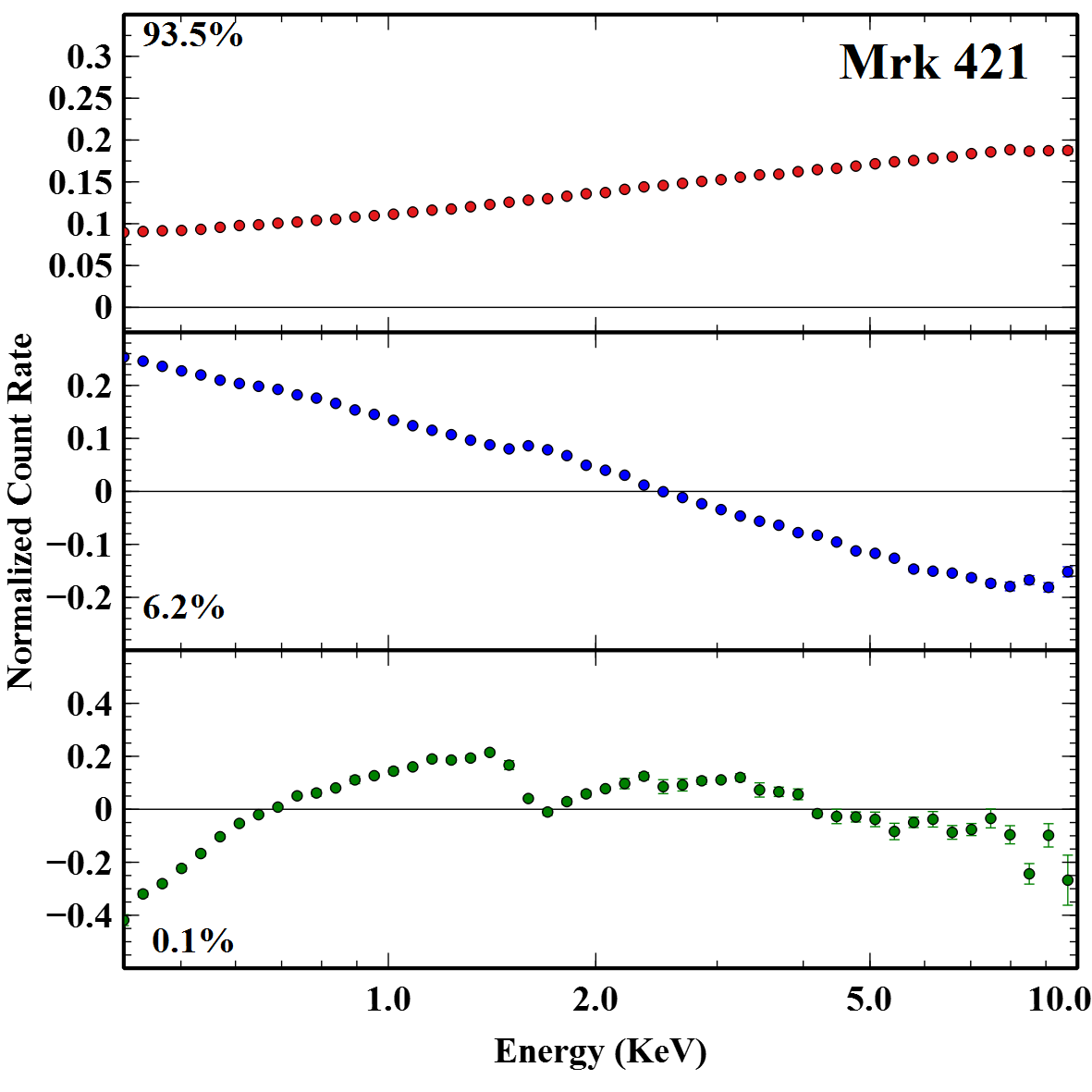}}
\end{minipage}
\hfill
\begin{minipage}[]{0.30\hsize}
\scalebox{0.40}{\includegraphics[angle=0]{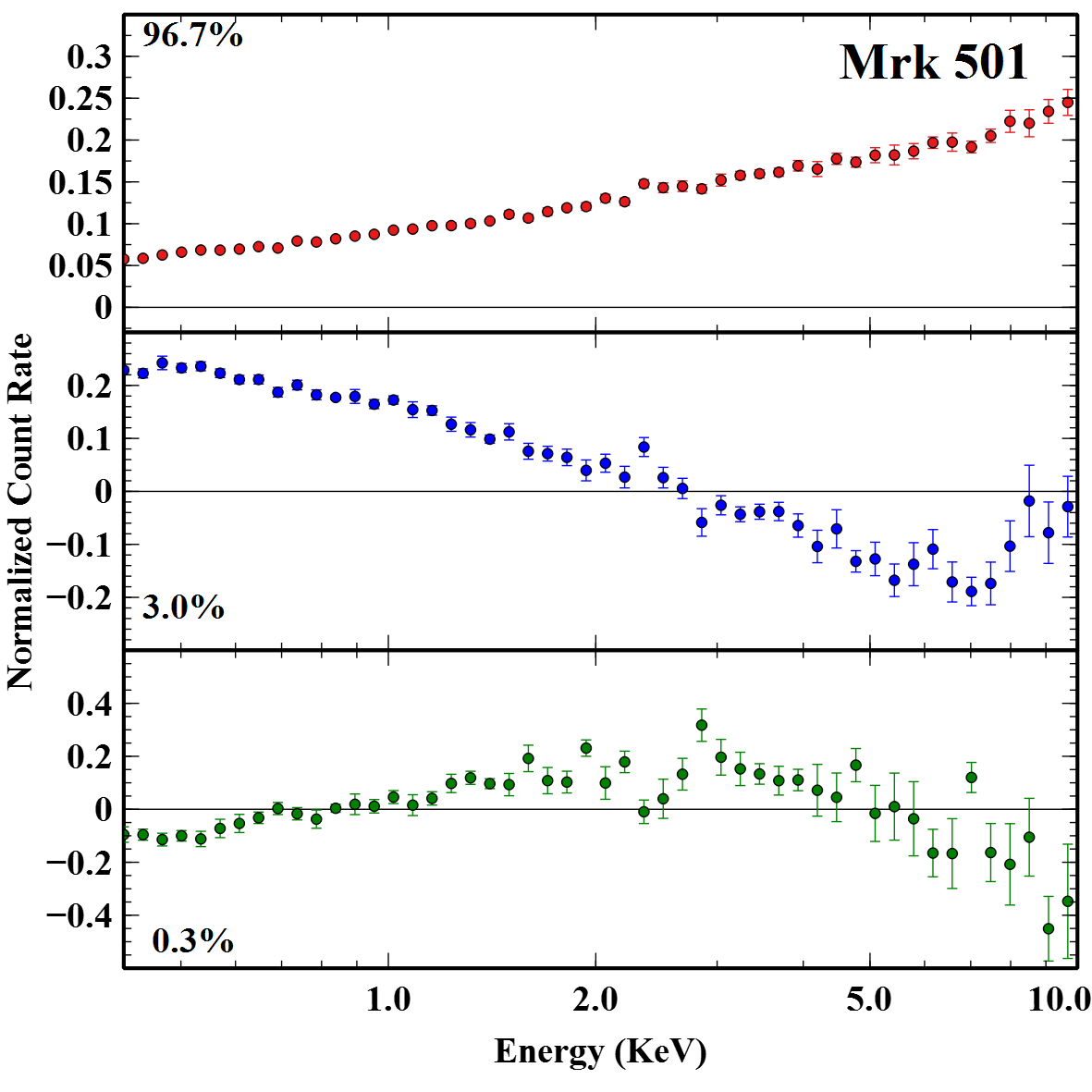}}
\end{minipage}
\hfill
\begin{minipage}[]{0.30\hsize}
\scalebox{0.40}{\includegraphics[angle=0]{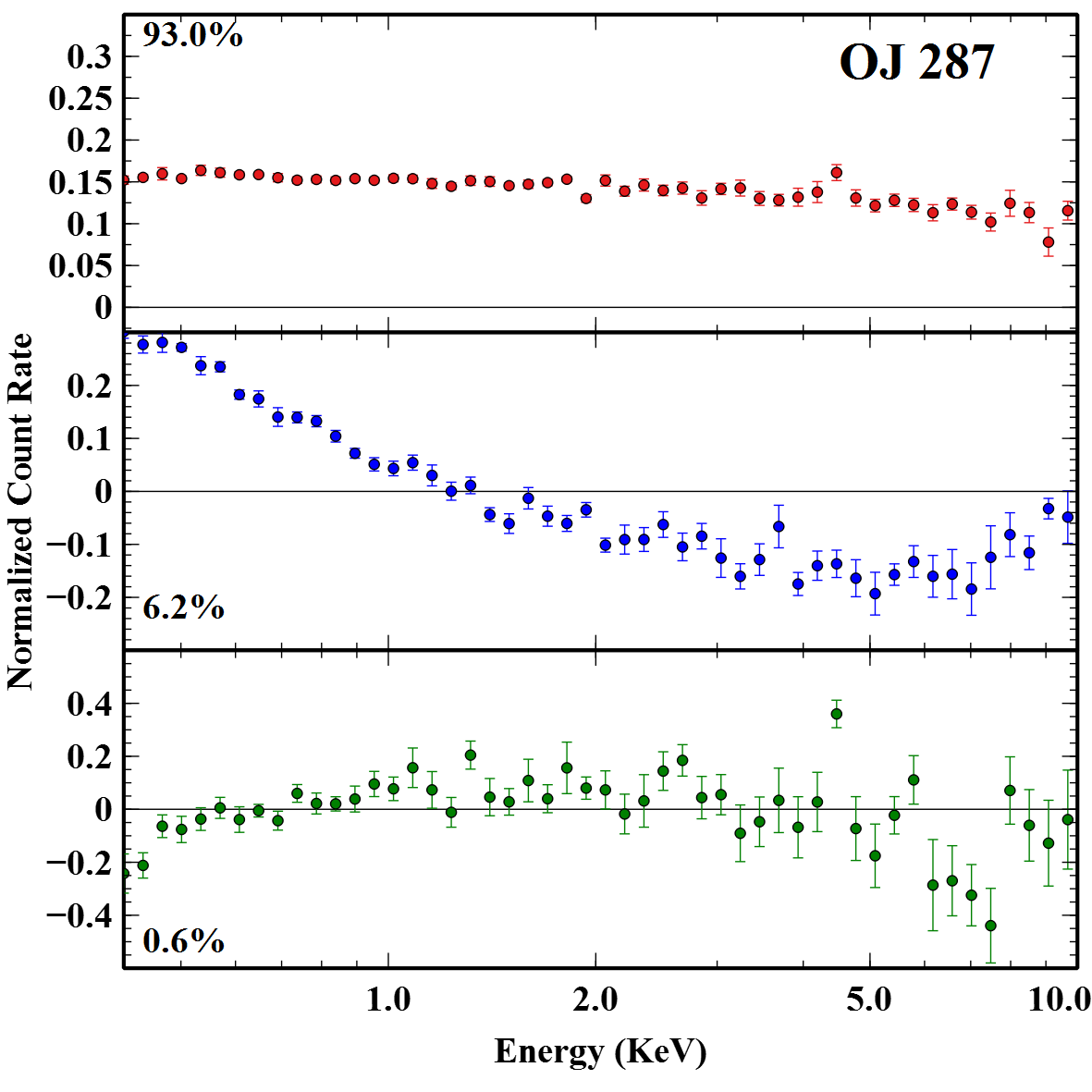}}
\end{minipage}
\hfill
\begin{minipage}[]{0.30\hsize}
\scalebox{0.40}{\includegraphics[angle=0]{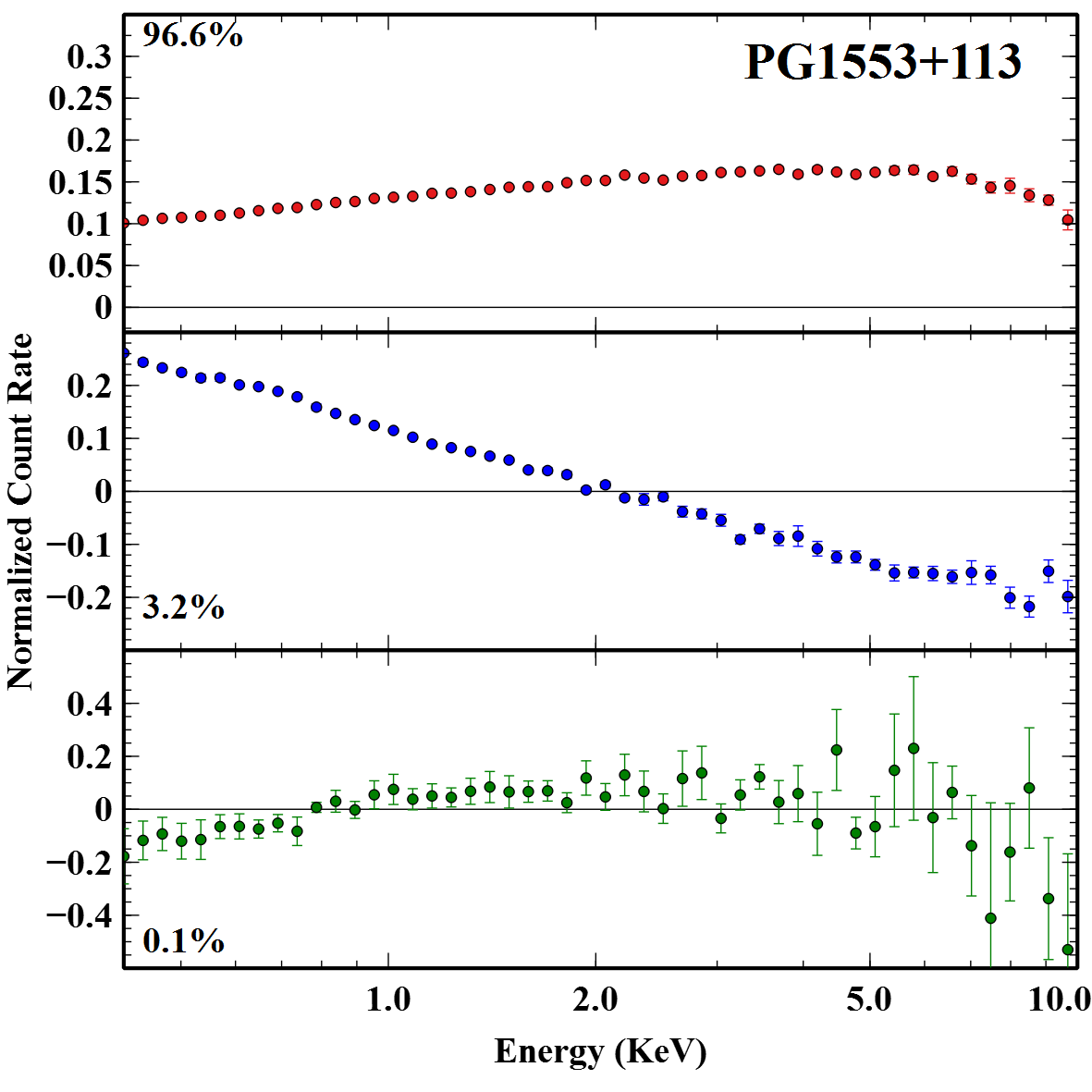}}
\end{minipage}
\hfill
\begin{minipage}[]{0.30\hsize}
\scalebox{0.40}{\includegraphics[angle=0]{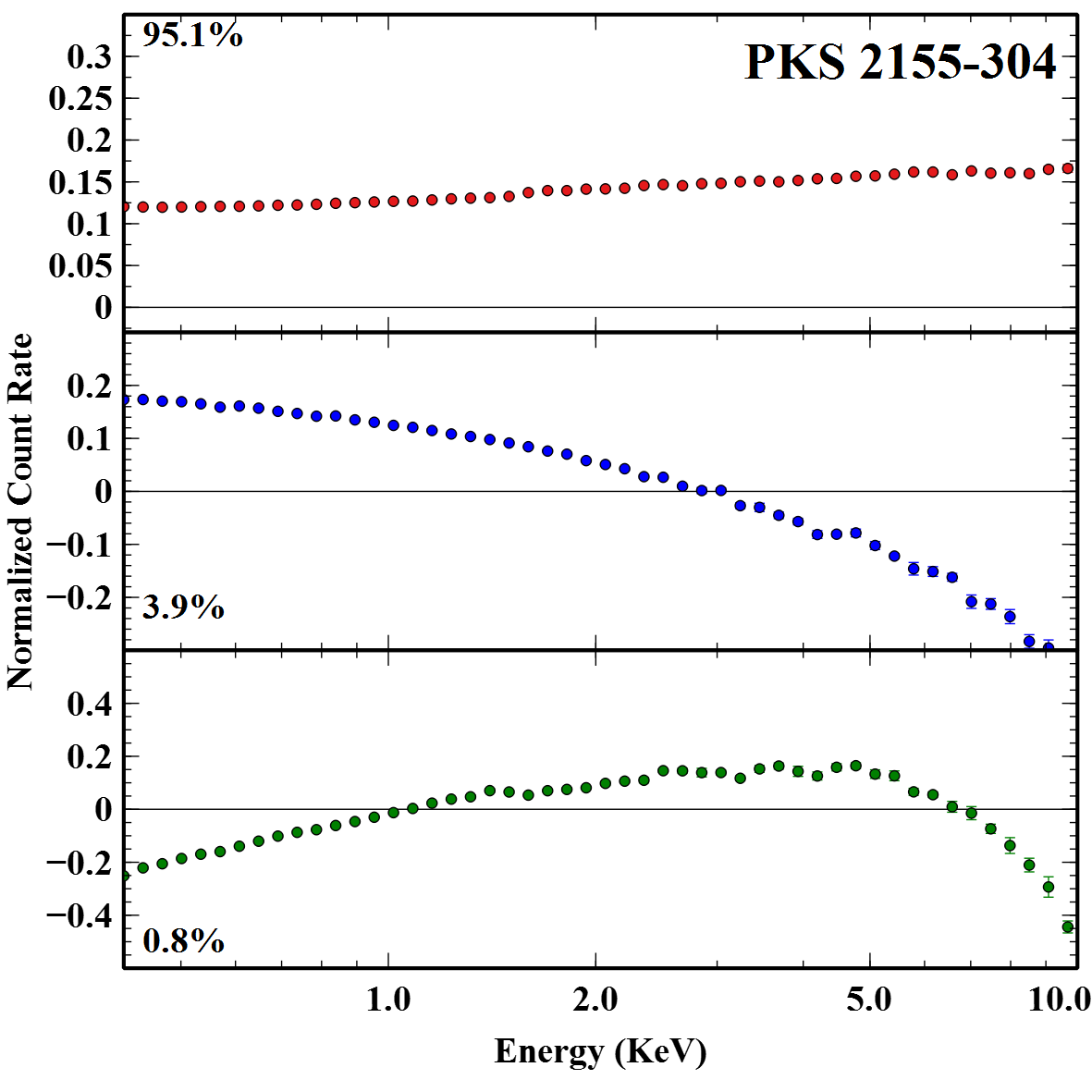}}
\end{minipage}
\hfill
\caption{Long-term PCAs for each of the seven objects that had at least four separate observations at different epochs. All show similar results. The first component is uniformly above zero and mostly flat, indicating changes in normalization. The second component shows an anti-correlation between low and high energies, consistent with changes in the photon index of a power law. The third component is shaped like an arch, and has no obvious physical explanation. Instead, it is likely a mathematical artefact of the PCA process caused by a breakdown of the linearity assumption (see Section 5). Components beyond the third were not significant in any object. These results indicate that long-term variability in blazars is dominated by changes in a power law model, varying both in shape and normalization.}
\end{figure*}

Figure 3 shows the results of the long-term PCA analysis for each object. The first three principal components are plotted in decreasing order of the fraction of the total variability that they are responsible for, expressed as a percentage in the plot. In every case, the remaining components showed no discernible shape and were not significant compared to the first three, and so are not plotted. \\

The results were very similar for every object, independent of the number or duration of observations, the time between observations, or the brightness of the object. In every case, the first principal component is uniformly above zero, meaning that all energy bands varied in a correlated manner. This shape is consistent with changes in the overall normalization of the spectrum. Changing the normalization of a given model causes the flux at all energies to rise or fall by the same amount, which explains the flat shape of this component. This interpretation of the first principal component is reinforced by simulations of power law variability presented in Parker et al (2015) and in Section 5 of this work. \\

The second component in each object shows an anti-correlation between flux changes in the low and high energy bands. This is consistent with a pivoting of the spectrum brought about by changes in the photon index of a power law model. \\

The third component has an arch-like shape, meaning energy changes in the low and high energies were correlated with each other and anti-correlated with changes in the energy band between them. This shape has no obvious physical explanation, and is probably a mathematical artefact of the PCA process. This is investigated further in Section 5.\\

These three components are consistent with a power law model varying in both normalization and $\Gamma$. The primary component is always due to changes in normalization for our sample. This accounts for most of the variability ($>$ 84 per cent) for our objects. The second component accounts for much less of the variability (3-15 per cent) and is attributed to pivoting of the power law. For our sample of blazars, the long-term X-ray variability over years appears to be dominated by changes in the brightness of the source, and less so by changes in the shape of the spectrum. \\

These power law components are to be expected of blazars, whose spectra are dominated by the effects of the jet, a highly-variable feature with a prominent power law shape. Note that this does not guarantee that a lone power law is sufficient to model the spectrum of any of these objects, only that the power law component is responsible for most of the variability. Due to the dominance of power law shapes in blazar spectra, and the close resemblance of these principal components to those produced by variations of a power law model, it is assumed that the vast majority of variability in blazars in the long term can be explained by changes in a single power law component. This does not mean that a more complicated model could not also work, but merely that a single power law is the simplest model that adequately explains the variability. Although these long-term PCAs are not identical, many of the differences between them can be explained without introducing a more complicated model by changing the way in which $\Gamma$ and normalization are assumed to be varying (see Section 5). \\

Additional comments on each individual PCA are presented in Appendix A.

\section{Short-term (single-observation) variability}
This analysis was performed on observations with at least 40ks of good time. These observations were split into 10ks parts, which comprised the input spectra for the PCA. The results show X-ray variability over timescales as short as a few hours. In many cases, only one component was significant, whereas there were always three significant components in the long-term analysis. \\

\begin{figure*}
\begin{minipage}[]{0.30\hsize}
\scalebox{0.40}{\includegraphics[angle=0]{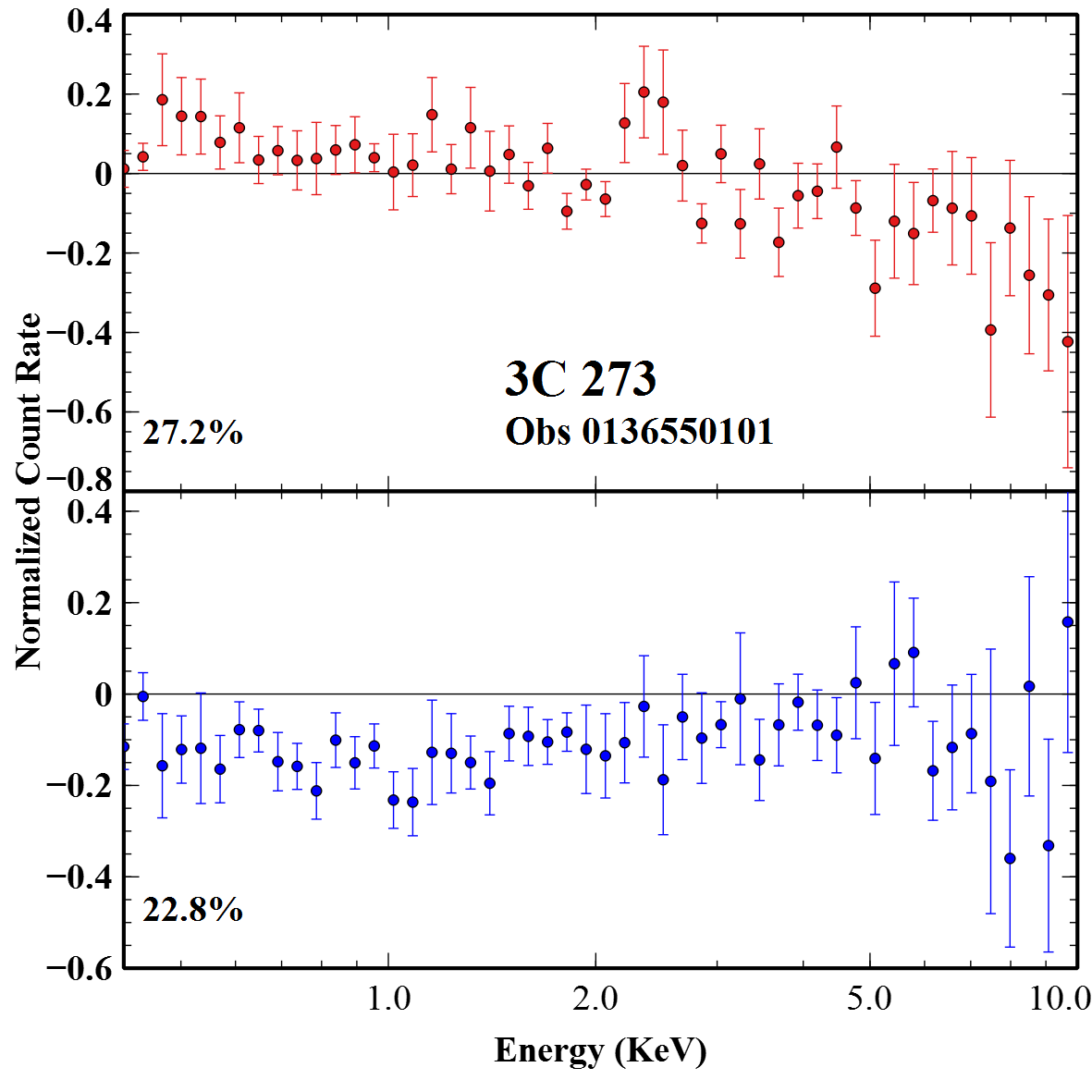}}
\end{minipage}
\hfill
\begin{minipage}[]{0.30\hsize}
\scalebox{0.40}{\includegraphics[angle=0]{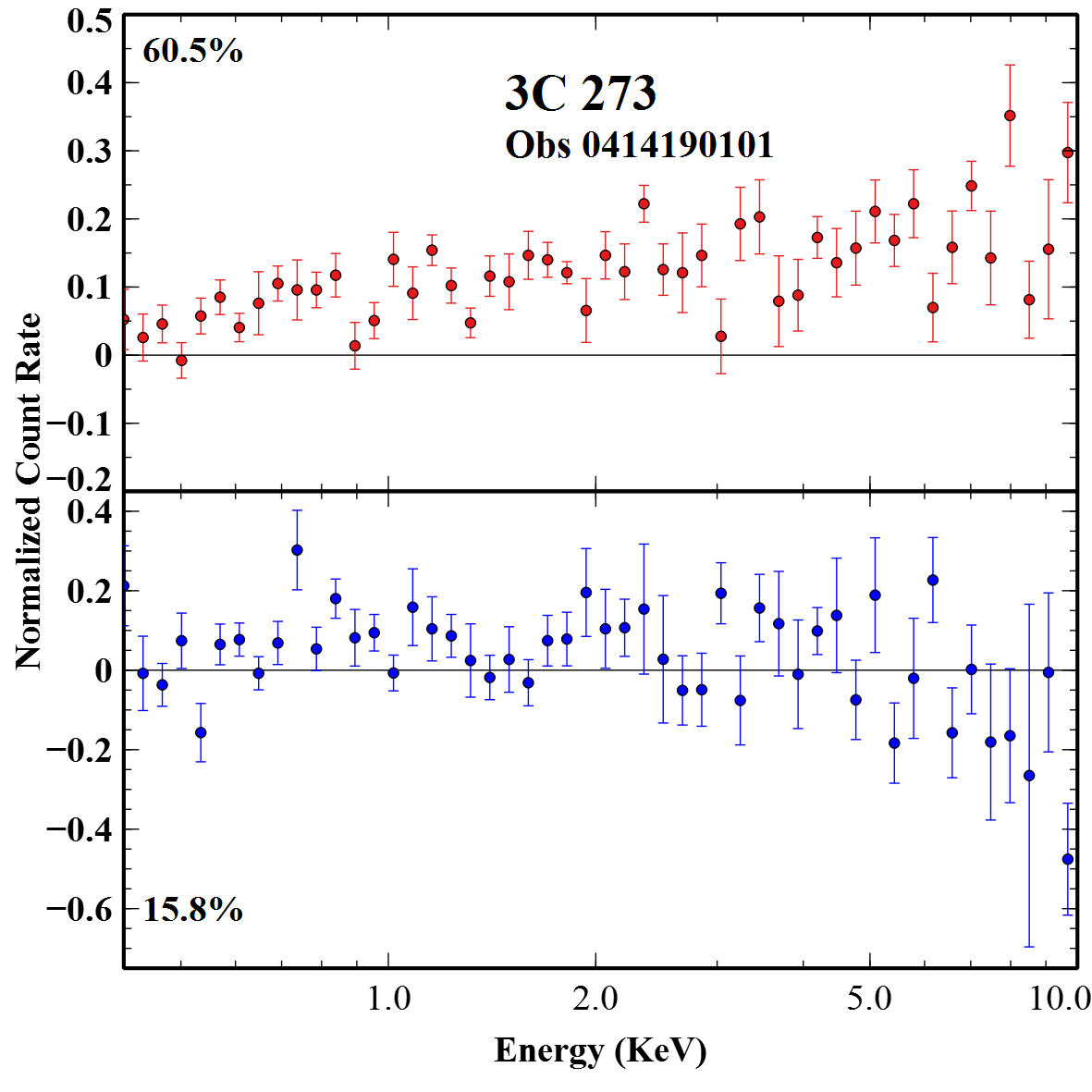}}
\end{minipage}
\hfill
\begin{minipage}[]{0.30\hsize}
\scalebox{0.40}{\includegraphics[angle=0]{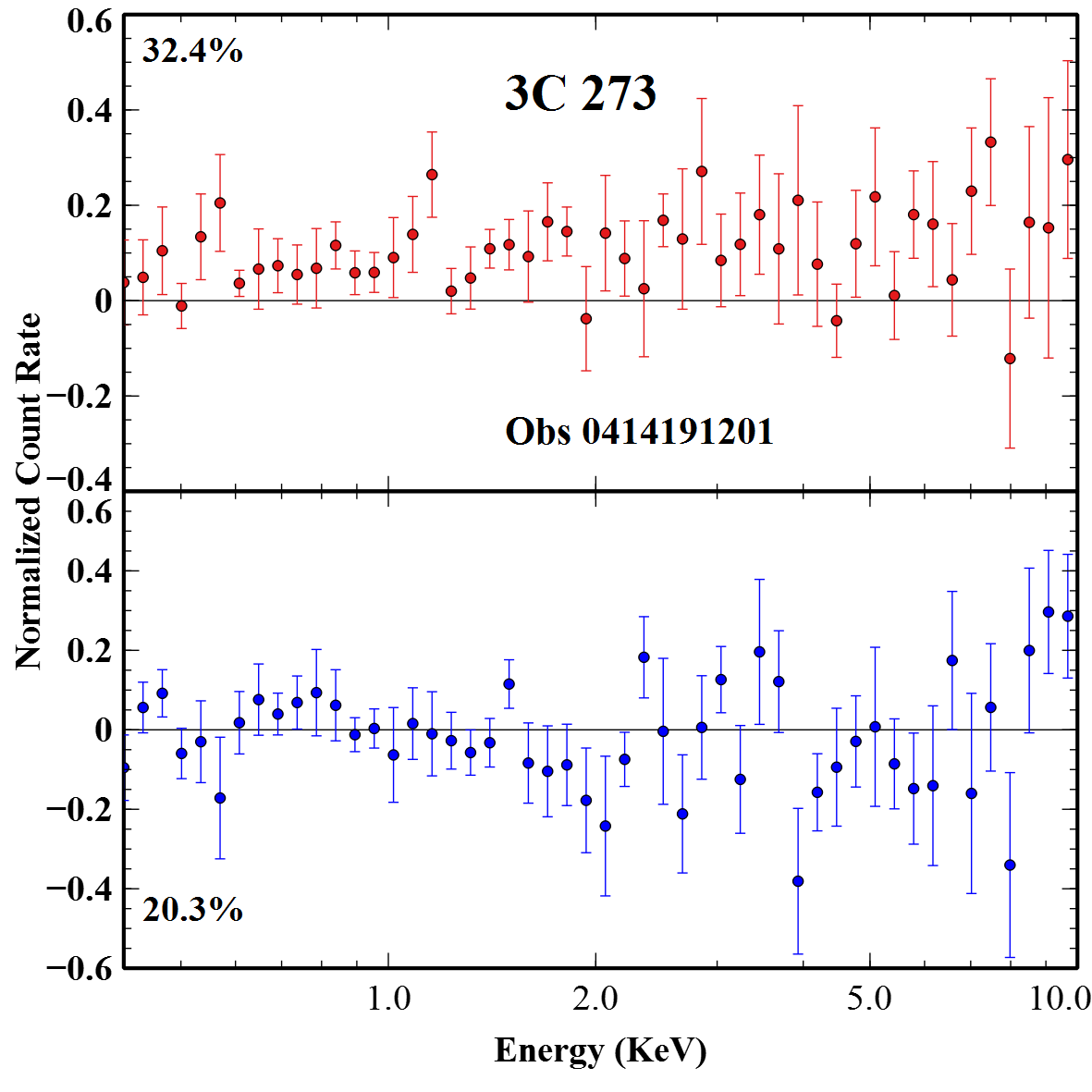}}
\end{minipage}
\hfill
\begin{minipage}[]{0.30\hsize}
\scalebox{0.40}{\includegraphics[angle=0]{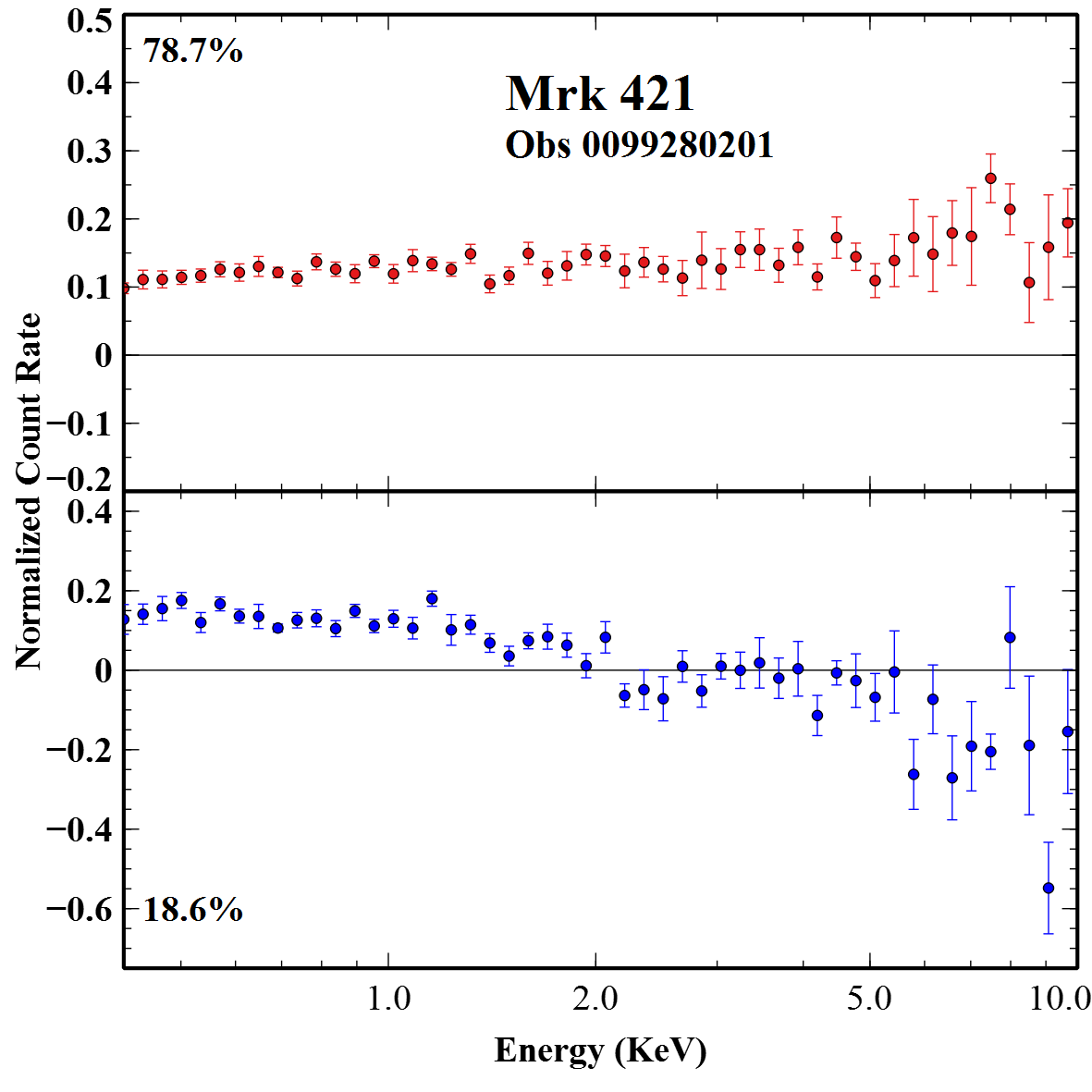}}
\end{minipage}
\hfill
\begin{minipage}[]{0.30\hsize}
\scalebox{0.40}{\includegraphics[angle=0]{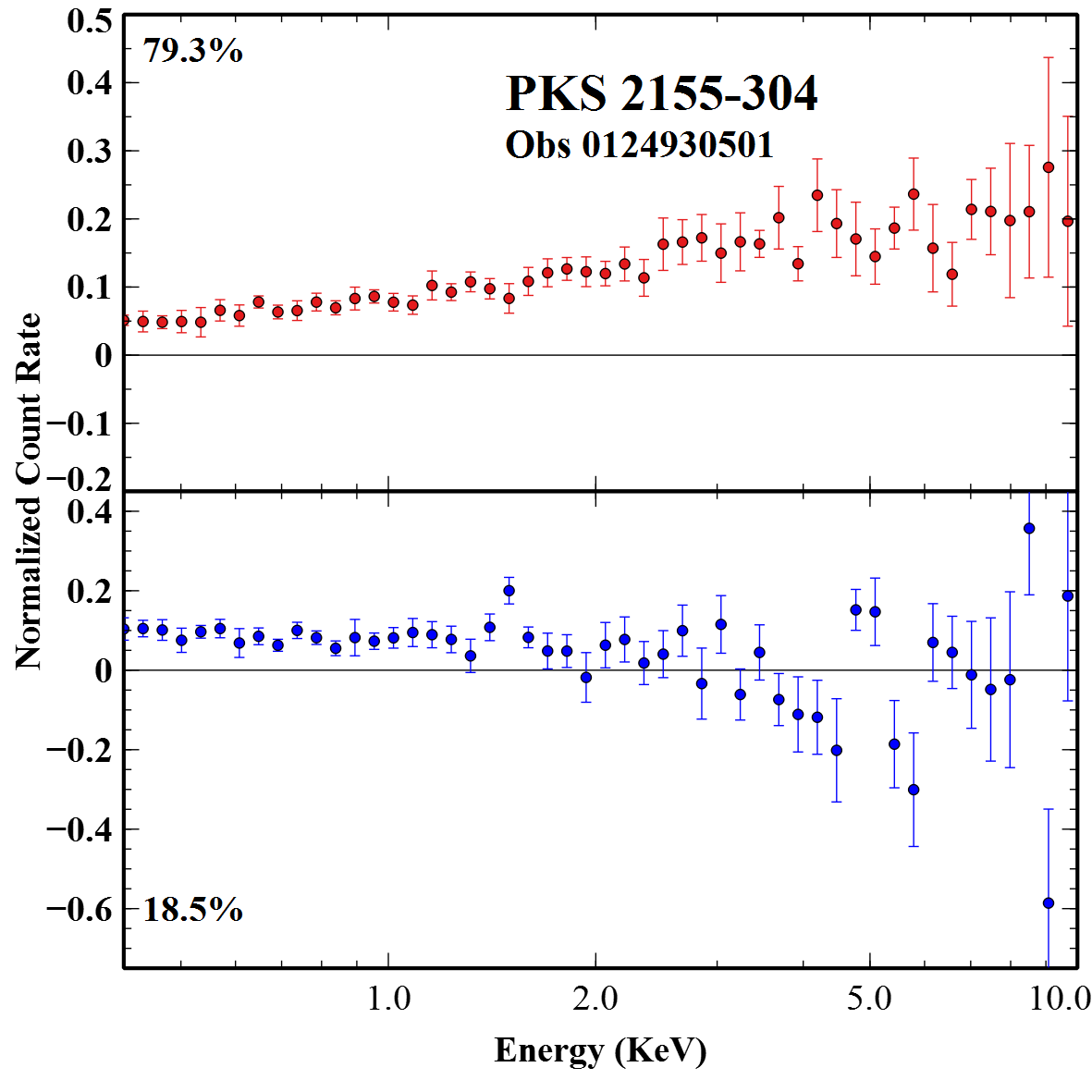}}
\end{minipage}
\hfill
\begin{minipage}[]{0.30\hsize}
\scalebox{0.40}{\includegraphics[angle=0]{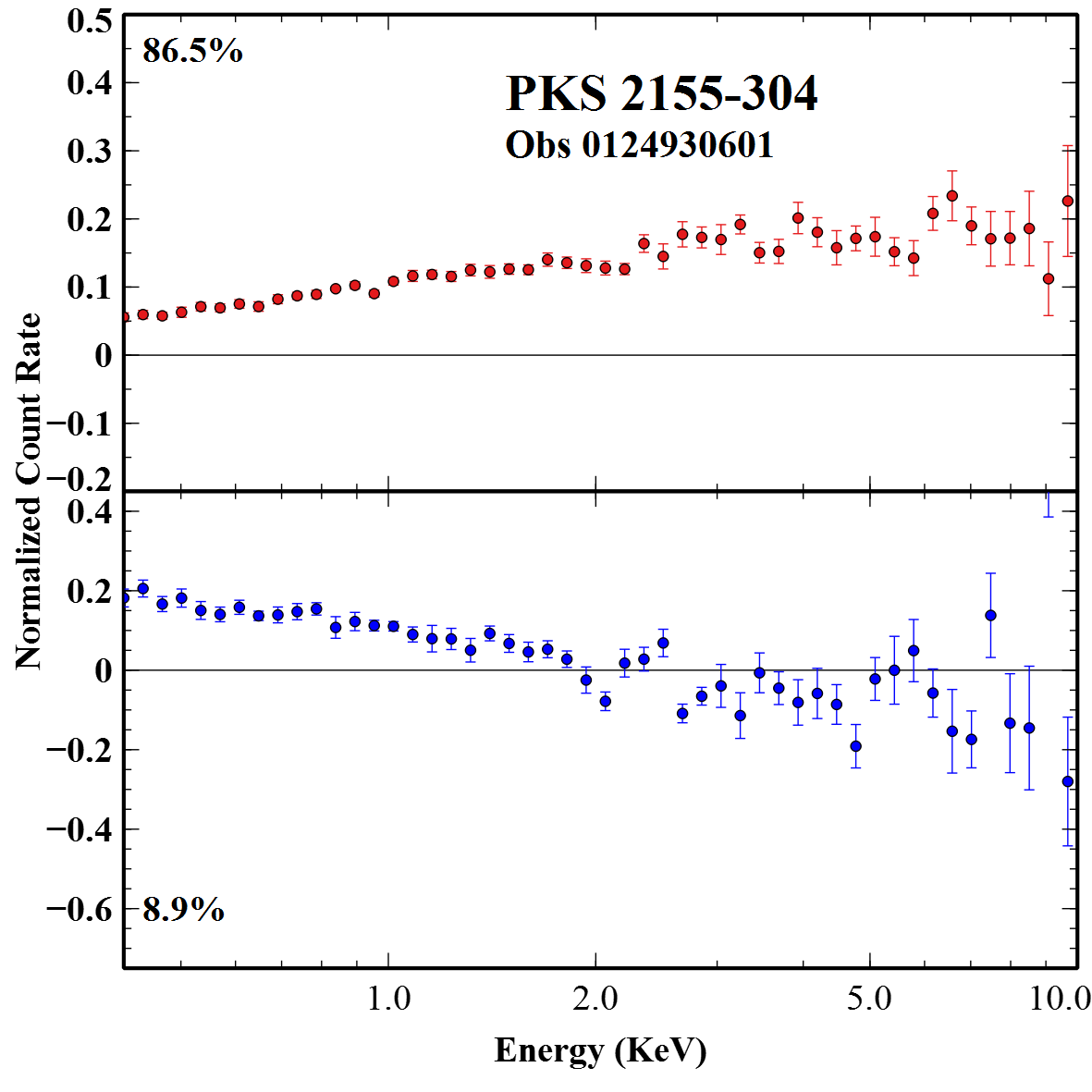}}
\end{minipage}
\hfill
\begin{minipage}[]{0.30\hsize}
\scalebox{0.40}{\includegraphics[angle=0]{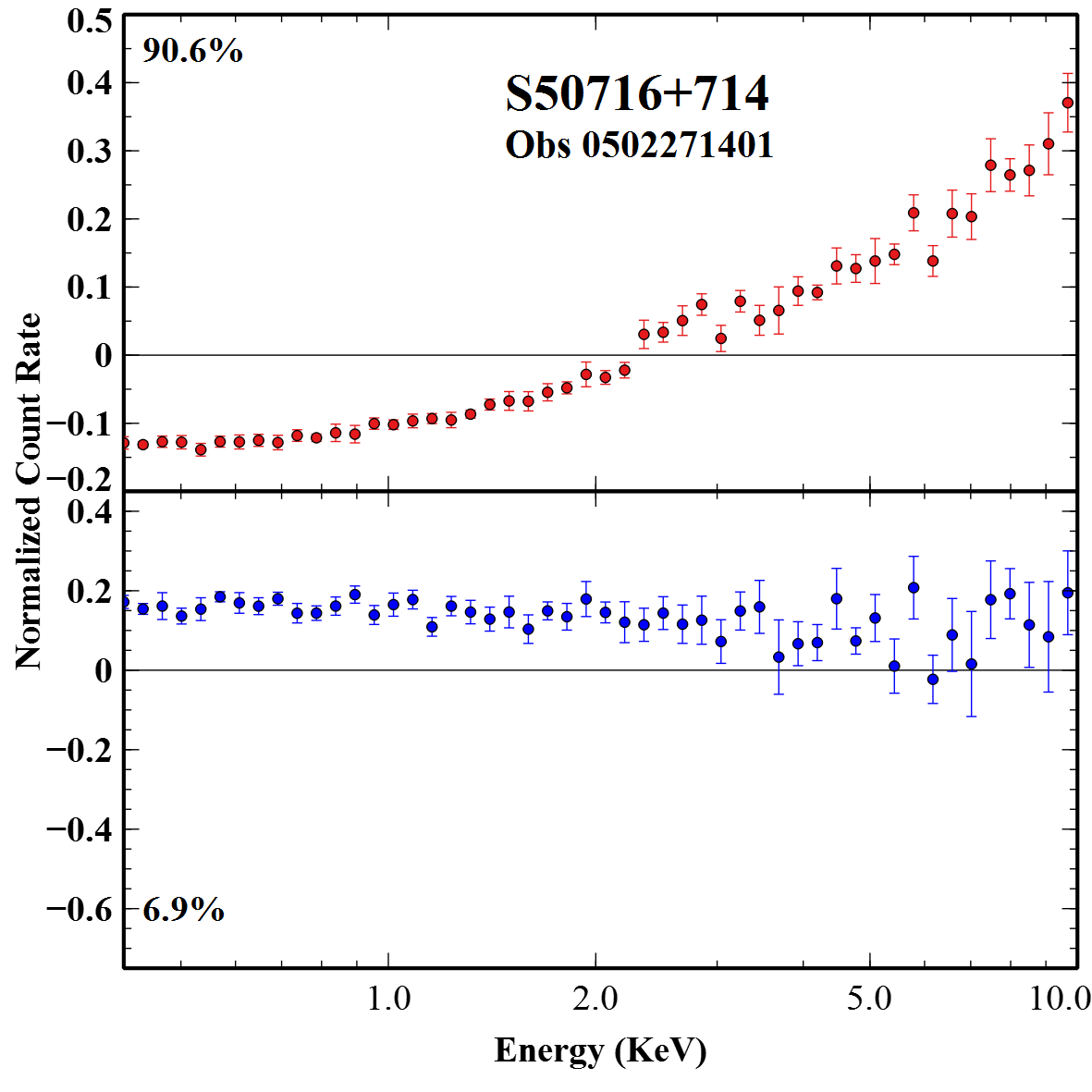}}
\end{minipage}
\hfill
\caption{Short-term PCAs that display components similar to those seen in the long-term analysis, corresponding to changes in the normalization and photon index of a power law. The results are not as pronounced as they are in the long-term case, which is likely a result of lower signal-to-noise.}
\end{figure*}

\begin{figure*}
\begin{minipage}[]{0.30\hsize}
\scalebox{0.40}{\includegraphics[angle=0]{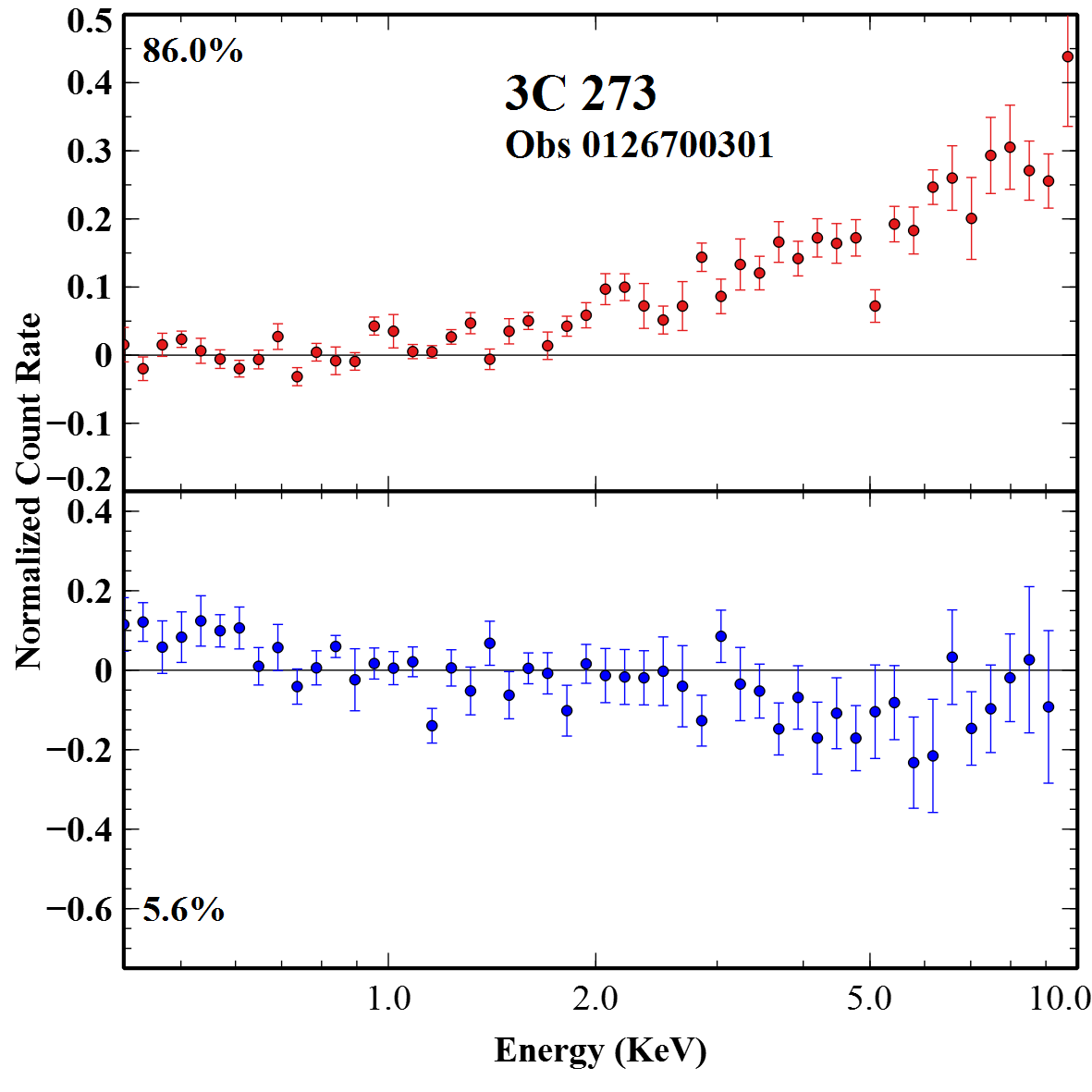}}
\end{minipage}
\hfill
\begin{minipage}[]{0.30\hsize}
\scalebox{0.40}{\includegraphics[angle=0]{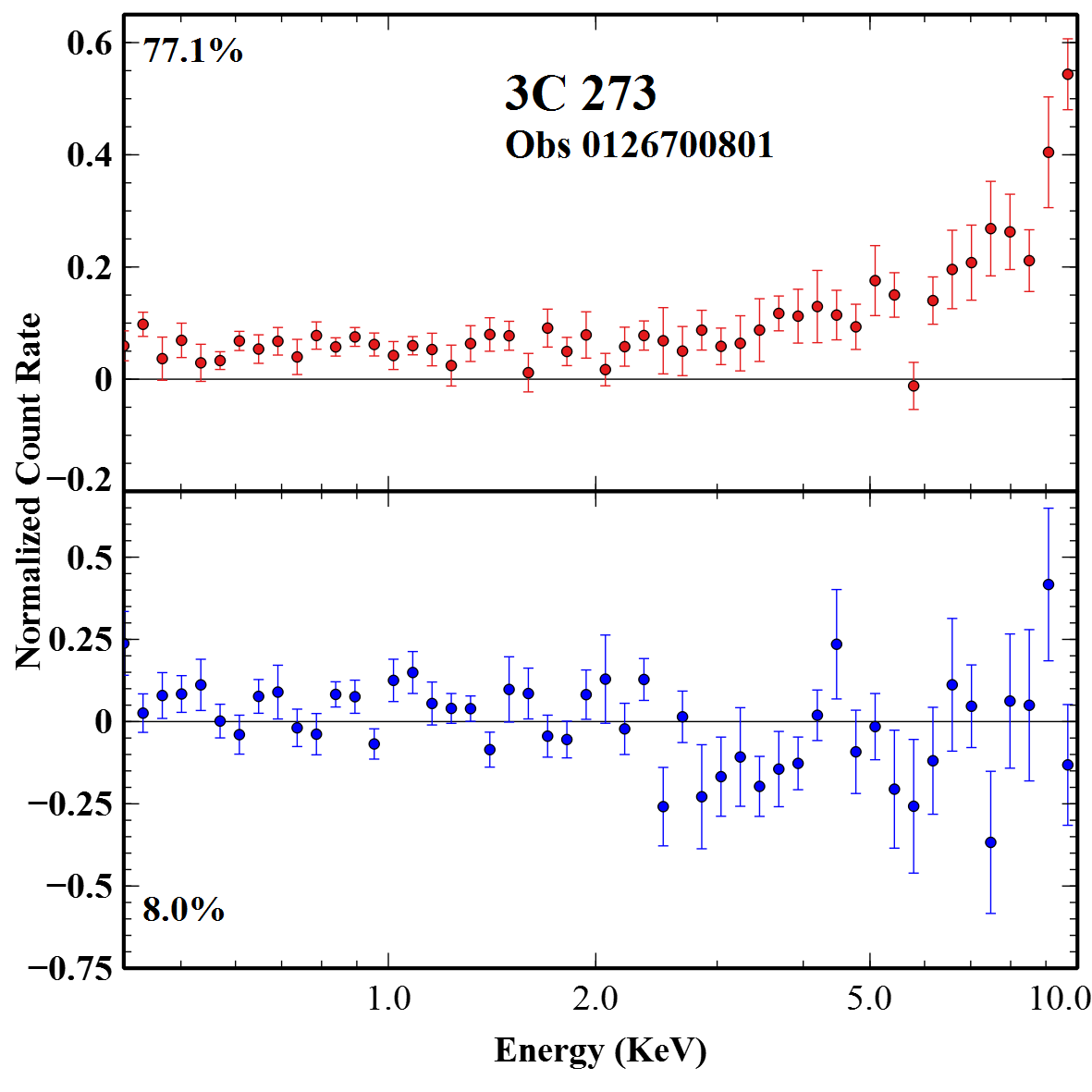}}
\end{minipage}
\hfill
\begin{minipage}[]{0.30\hsize}
\scalebox{0.40}{\includegraphics[angle=0]{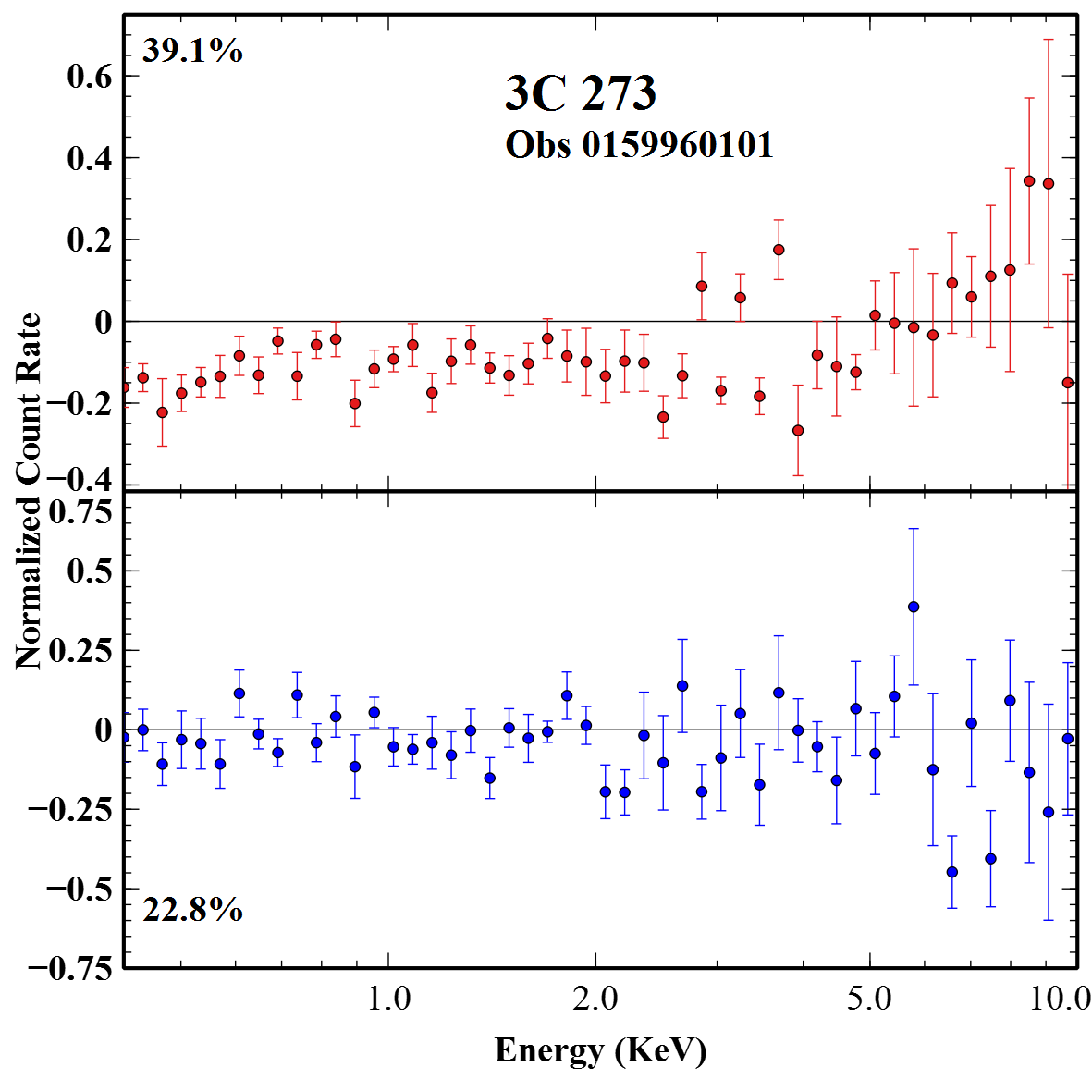}}
\end{minipage}
\hfill
\begin{minipage}[]{0.30\hsize}
\scalebox{0.40}{\includegraphics[angle=0]{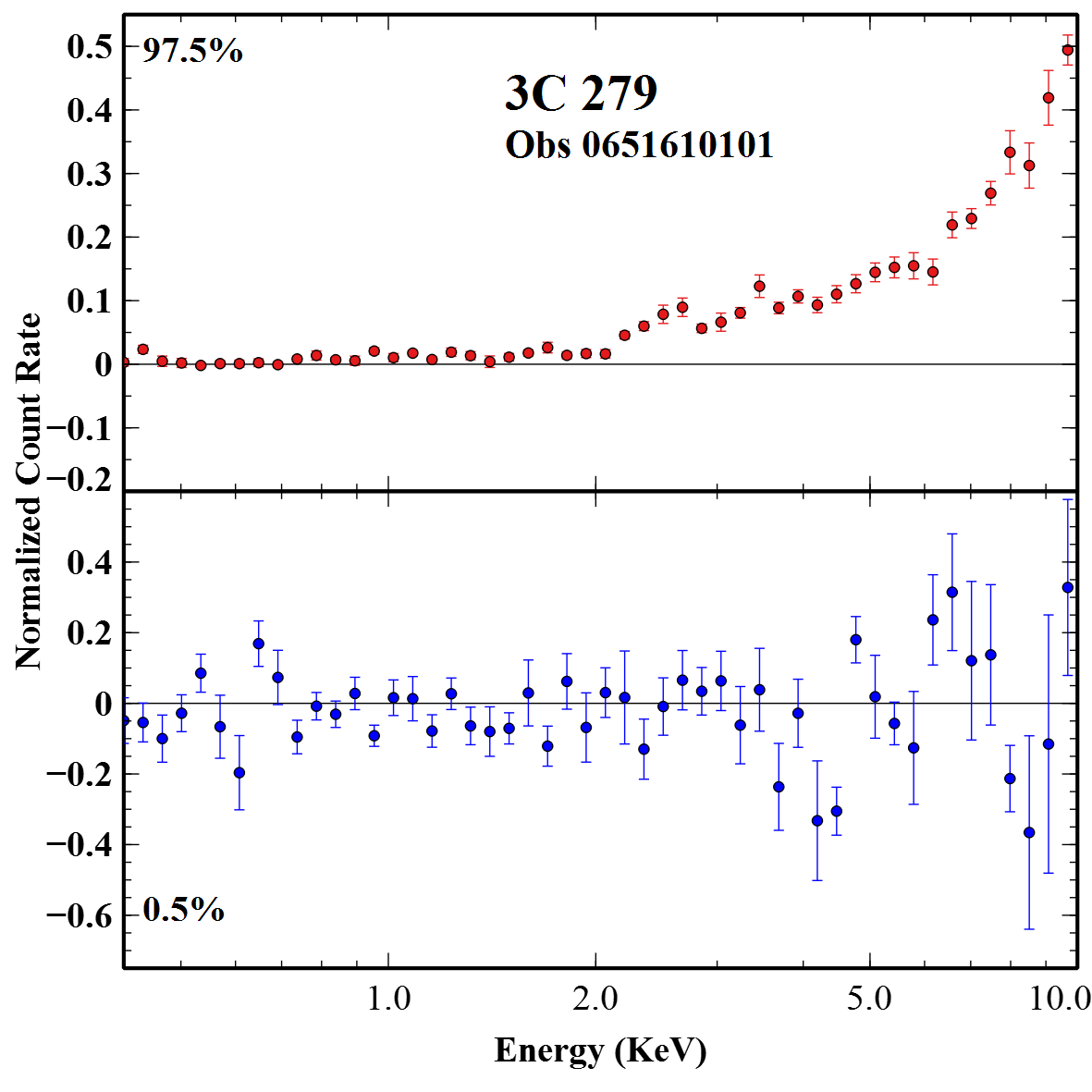}}
\end{minipage}
\hfill
\begin{minipage}[]{0.30\hsize}
\scalebox{0.40}{\includegraphics[angle=0]{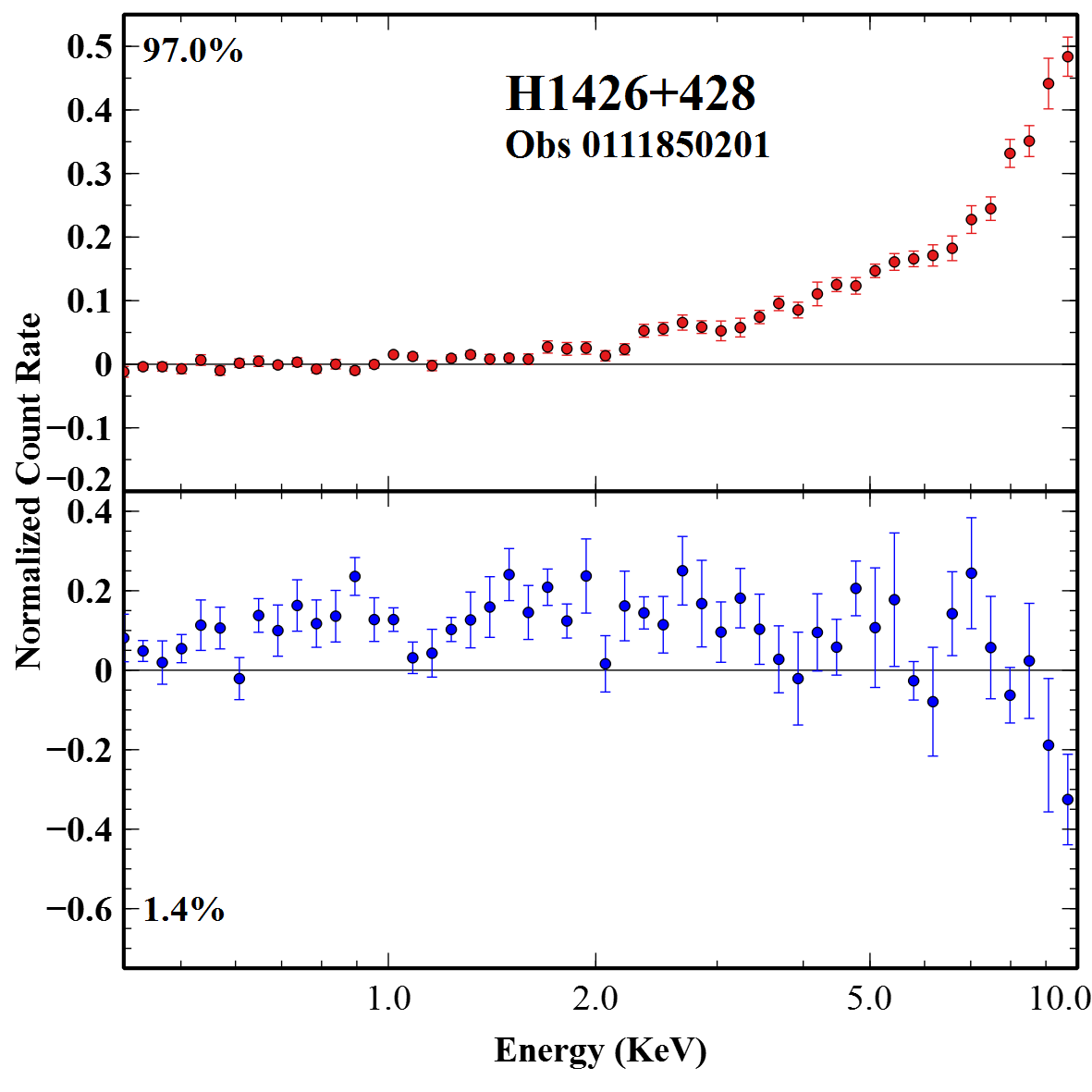}}
\end{minipage}
\hfill
\begin{minipage}[]{0.30\hsize}
\scalebox{0.40}{\includegraphics[angle=0]{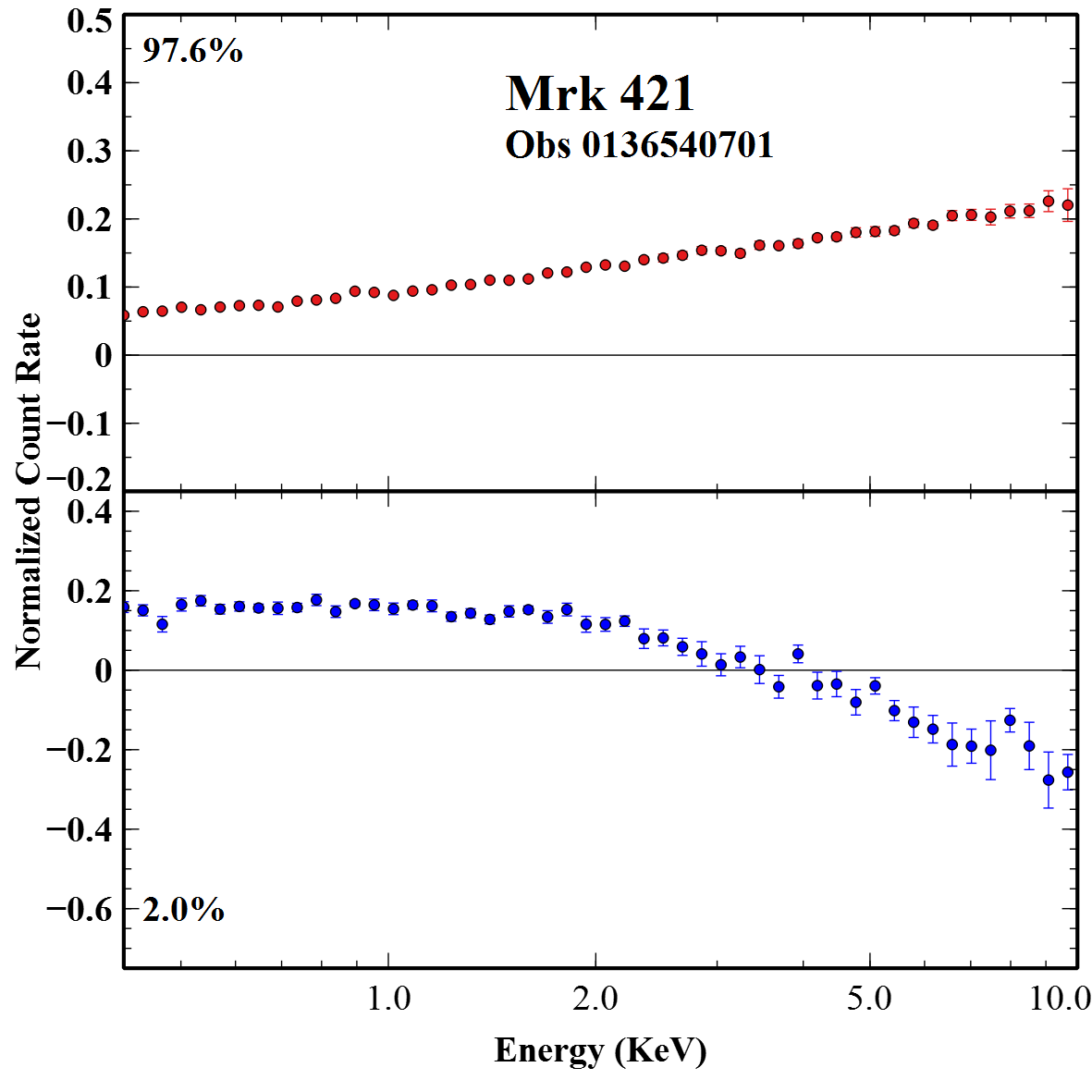}}
\end{minipage}
\hfill
\begin{minipage}[]{0.30\hsize}
\scalebox{0.40}{\includegraphics[angle=0]{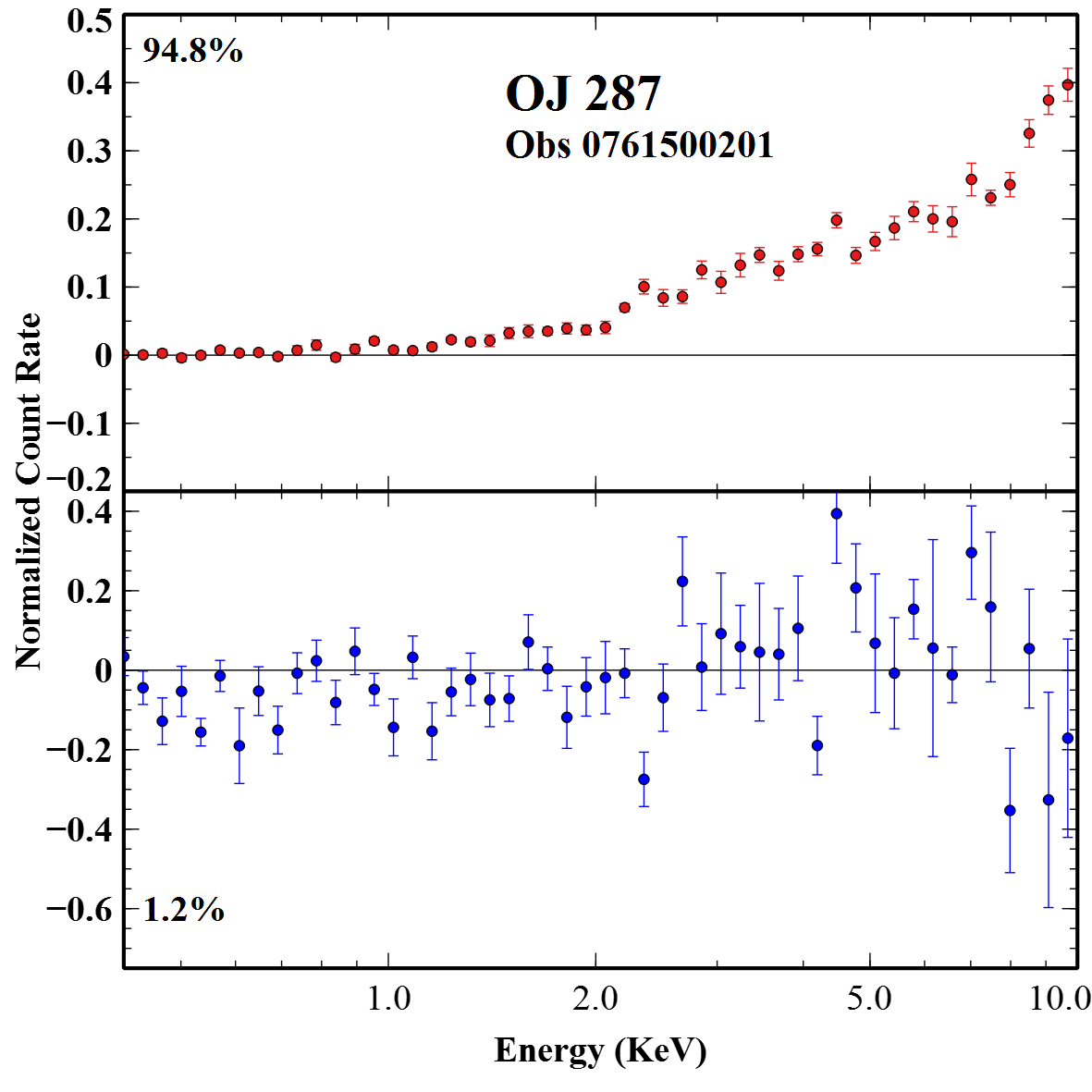}}
\end{minipage}
\hfill
\caption{Short-term PCAs with shapes unlike those seen in the long-term analysis.}
\end{figure*}
There were several observations that showed no discernible shape in any component, indicating that the object was either constant at that time, or varying on timescales longer than the observation itself. These are not plotted. \\

The remaining short-term PCAs can be divided into two groups: those with one or more components similar to those seen in the long-term analysis, presented in Figure 4, and those showing shapes unique to the short-term analysis, shown in Figure 5. \\

The first category includes PCAs that display normalization (flatter, uniformly above zero) and/or $\Gamma$ (pivoting) components similar to those seen in the long-term (Figure 3). There are still some differences between the two timescales, however. Some observations show only one noticeable component, and not both, such as 3C 273 observation 0414190101, which lacks a clear pivoting component. Others have the pivoting component as more significant than the normalization component, with S50716+714 being the best example. In every case, the third component seen in the long-term results does not appear with any significance in the short-term. This is most likely due to lower signal-to-noise in the short-term.\\

Overall, these results are indicative of the same sorts of changes seen over long timescales, and the long-term results could be seen as the summation of many years worth of these short-term changes. \\

The second group exhibits at least one component that differs from the straightforward normalization and pivoting components seen elsewhere. For example, the first component of 3C 273 observation 0126700301 and the second component of Mrk 421 observation 0136540701 display a broken power law shape. They have no variability below some break point (around 2 keV) and then increasing variability with increasing energy after the break. Some of the components in this group show an additional upward curve, displaying an almost exponential shape after the break. The first components of H1426+428 observation 011850201 and OJ 287 observation 0761500201 are good examples of this shape. The broken power law shapes seen in Figure 5, but not in the long-term analysis, would seem to indicate an intrinsic difference between long and short-term flares in blazars. Not being seen in the long-term PCAs means they are insignificant over long timescales. It is unclear what, if anything, these components correspond to physically, and simulations have been unable to replicate the upward-curving shape. To ensure that these shapes were not influenced by background effects, some short-term PCAs were performed again with background subtraction turned off. This did not significantly effect the results beyond introducing more noise, indicating that these shapes are part of the source spectrum. \\

We also note that some objects fall in both groups. For example, 3C 273 sometimes exhibits rapid variability consistent with the long-term (yearly) variations, while also having epochs where the PCA spectral shape is unique. \\

Additional comments on each observation's PCA are presented in Appendix B.\\

\section{Models}

The model-independent nature of PCA is useful, but it has its drawbacks. One weakness is that it is often unclear what the resultant principal components correspond to physically. Simulations can help with this. In this Section, PCA is performed on a set of fake spectra generated according to a given model, with each spectrum varying the model parameters randomly within a certain range. Performing PCA on a known model allows for comparisons to the real data to be made and can identify how shapes are associated with  physical parameters. As one would expect, simulations of a power law varying in both normalization and $\Gamma$ can closely reproduce the results of the long-term PCAs presented in Section 3 (Parker et al 2015). Furthermore, many of the differences between the various PCAs can be reproduced by changing the ways in which the model parameters vary in relation to each other.  \\

Figure 6 shows the results of PCAs performed on three sets of 100 simulated spectra conforming to a power law model varying in both normalization and $\Gamma$. In the first PCA, $\Gamma$ varied randomly by up to 10 per cent, while normalization varied randomly by up to a factor of four. These amounts were determined through trial and error for the purposes of reproducing the results as closely as possible while remaining within reasonable ranges for real objects. As seen in the long-term PCAs, there are three significant components: a flat component representing changes in normalization, a pivoting component representing the changing slope of the power law, and an arch-shaped third component. These are the archetypal power law results that explain most of the shape in the long-term PCAs. \\

The second simulation put more emphasis on $\Gamma$, allowing it to vary by up to 25 per cent. This induces an upward slope on the first component, which is seen in several of our objects (H1426+428, Mrk 421, Mrk 501, and PKS 2155-304). Increasing the amount $\Gamma$ varies increases this slope of the first principal component. The final simulation in Figure 6 varied the parameters as in the first simulation, except they now varied in a correlated manner, rather than independently. Changes were correlated such that normalization increased or decreased as $\Gamma$ increased or decreased, resulting in a softening of the spectra as their brightness increased. This induces a negative slope in the first component, as well as weakening the second component slightly at high energies, and suppressing the third component entirely. All three of these effects are seen to some degree in OJ 287. \\

While simulations can do a good job of reproducing PCA shapes, each component's fractional contribution to the total variability is less easy to simulate. In simulations of a varying power law, the first component generally accounts for  $>$90 per cent of the variability, whereas the third component is responsible for only a tiny fraction, even compared to the results from real data. Because of the presence of noise in the real data, and the fact that the noise contributes significantly to the total variability, it is difficult to reproduce the correct share of the total variability for each component. \\

\begin{figure*}
\begin{minipage}[]{0.30\hsize}
\scalebox{0.40}{\includegraphics[angle=0]{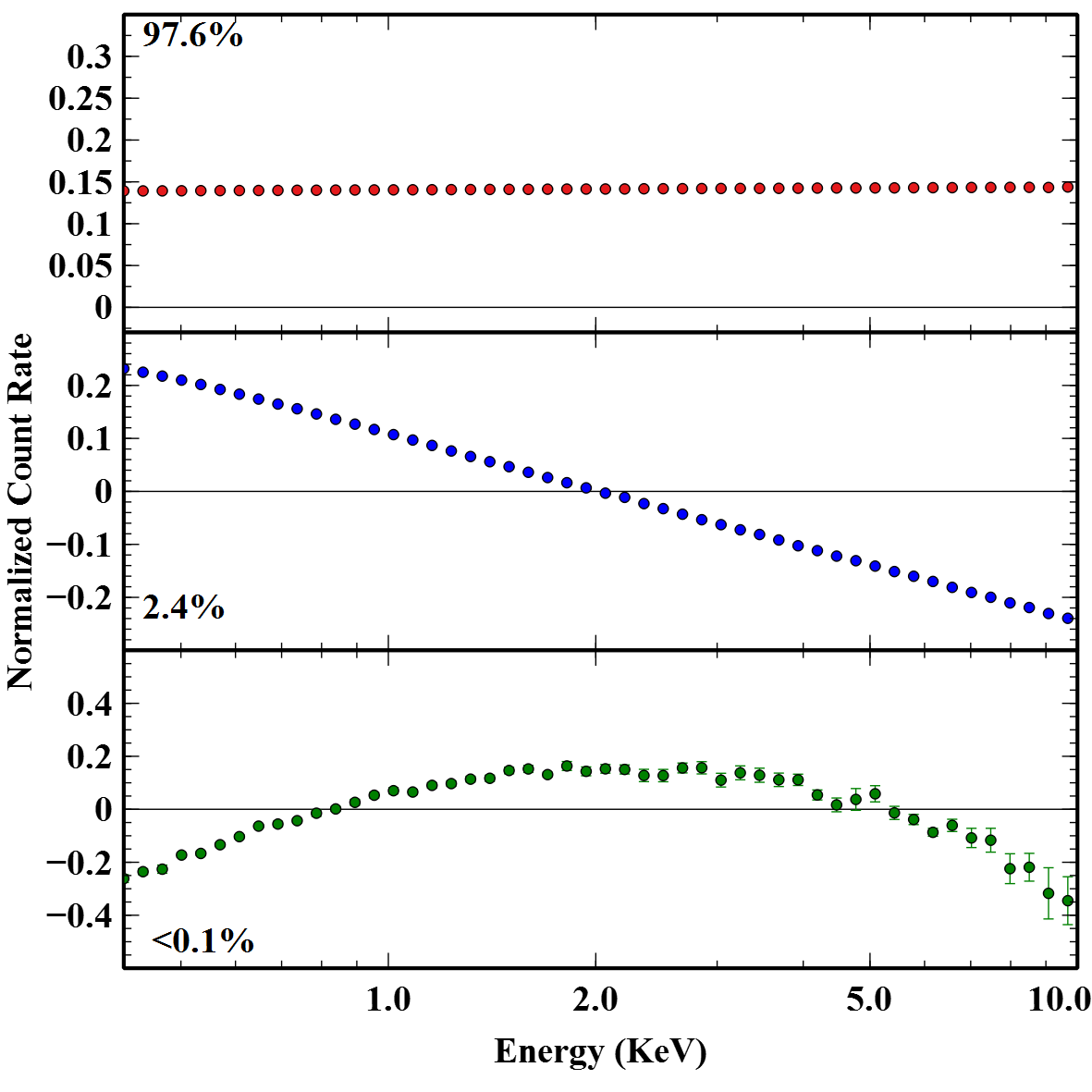}}
\end{minipage}
\hfill
\begin{minipage}[]{0.30\hsize}
\scalebox{0.40}{\includegraphics[angle=0]{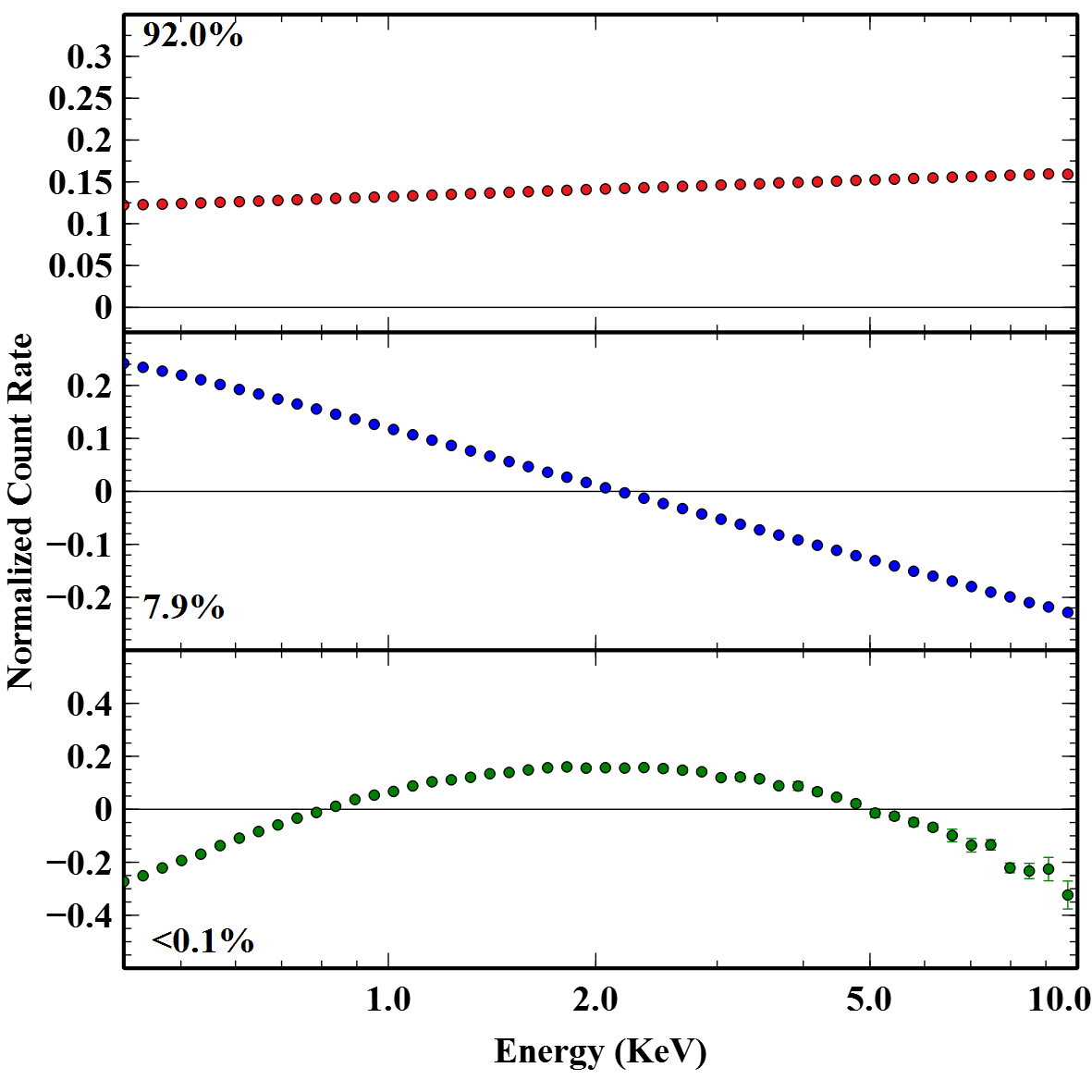}}
\end{minipage}
\hfill
\begin{minipage}[]{0.30\hsize}
\scalebox{0.40}{\includegraphics[angle=0]{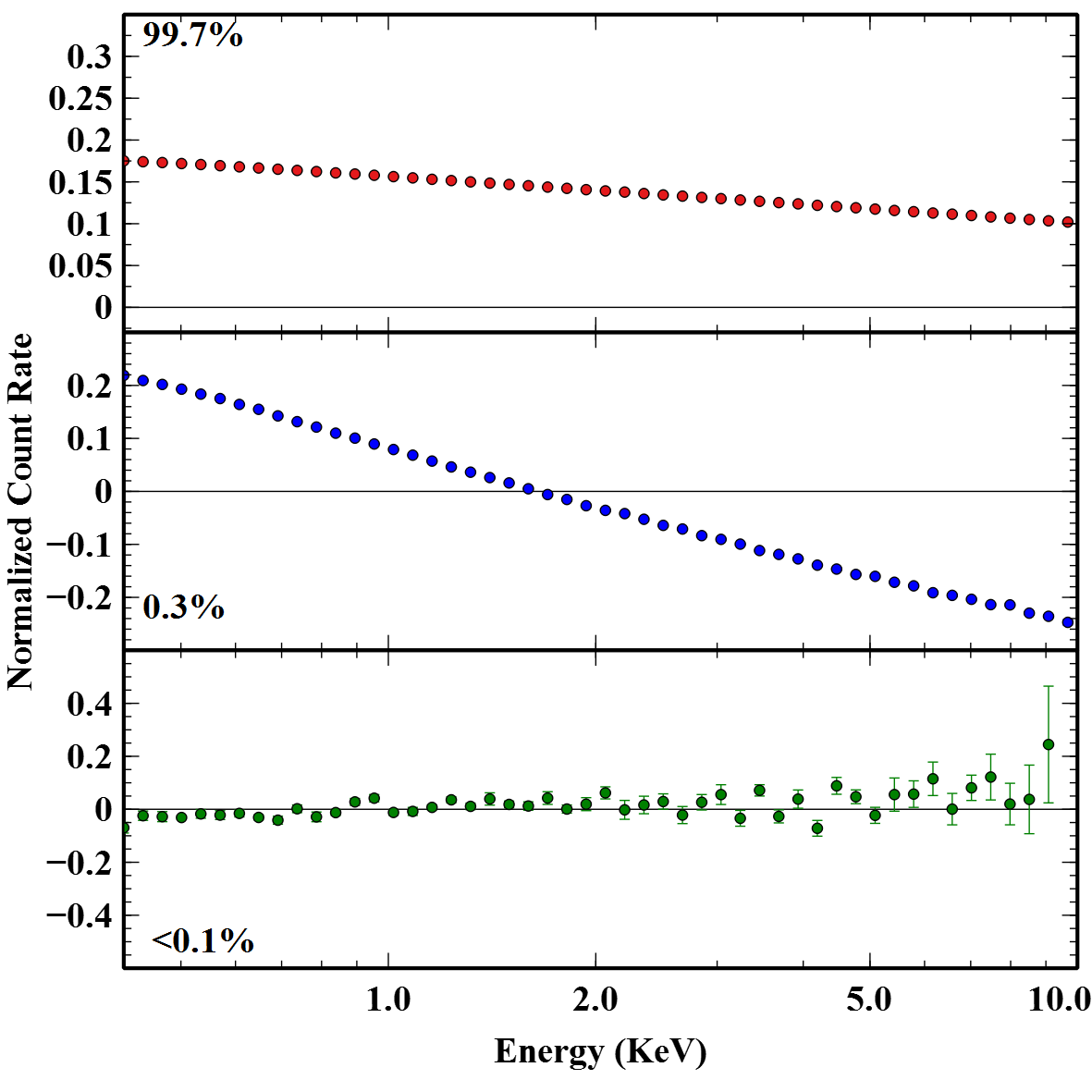}}
\end{minipage}
\hfill
\caption{Results of PCA performed on 100 simulated power law spectra varying in normalization and 
$\Gamma$. Left: $\Gamma$ varied randomly by up to 10 per cent, normalization varied by up to a factor of four. Middle: $\Gamma$ varied by up to 25 per cent, while normalization still varied by up to a factor of four. Right: Same as in the left panel, except the variation in both parameters were correlated; normalization increased or decreased as $\Gamma$ increased or decreased, leading to a softening of the spectra with brightness.}
\end{figure*}

Notably, the third principal component still appears in simulated PCAs, even though we can be absolutely certain that the model can be described by only two components. This indicates that the third principal component does not correspond to any sort of model parameter, but rather is created as a by product of the PCA process. One of the assumptions made during PCA is \textit{linearity}, meaning that the new basis vectors are a linear combination of the old ones, and that any correlations among the original data set are linear. However, this is not entirely true for most spectra, even those conforming to a simple power law model. A power law changing its slope, for example, can not be described linearly. A linear approximation of a power law will always undershoot the model at both low and high energies, and overshoot it in the middle, no matter the slope of the power law. The PCA process sees this as a problem, and fixes it by creating a new component with just the right shape to make up for the places that the linear approximation fails. \\

Since the second (pivoting) component is the one responsible for describing the changing shape of the power law, this third component should be strongest whenever the second component is strongest (in either direction). Strongest, in this case, refers to the normalizations of the principal components. If this explanation is true, a plot of the second component against the third should display a V-shape. \\

Figure 7 shows the normalizations of the second and third components plotted against each other for PCA performed on 1000 simulated power law spectra that varied as in the second simulation in Figure 6. The results show that the third component grows in strength as the second component increases in either direction, as expected. This is how the PCA code accounts for its inability to describe a changing power law using a linear function. It also explains why the third simulation in Figure 6 shows a suppressed third component: by varying $\Gamma$ and normalization together, we have introduced linear correlation in the data, therefore reducing the need for a corrective third component. Even though this third component is not physical, it does appear in real data and thus can still be used as an indicator of a changing power law. The same can not be said for the short-term PCAs, however. None of the short-term PCAs had a third significant component, due to reduced signal-to-noise. Non-linearities within the data would certainly have some effect in the short term, but that effect is too small to detect in our sample. 

\begin{figure}
\begin{center}
\begin{minipage}{0.99\linewidth}
\scalebox{0.60}{\includegraphics[angle=0]{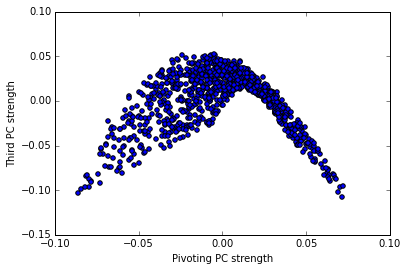}}
\end{minipage}  \hfill
\end{center}
\caption{Normalizations of second and third principal components plotted against each other for PCA of 1000 simulated power law spectra. The third component is strongest wherever the second component is strong in either direction. In other words, wherever a linear approximation would differ significantly from a power law, the third component accounts for the difference.}
\end{figure}

\section{Discussion}
The PCA of blazar X-ray variability over years indicates the variation arising from changes in a power law component. The primary principal component for all sources in our sample indicated that changes in brightness (normalization) is the dominant factor, responsible for $>$84 per cent of the variability in each source. The secondary effect (component two) was the changes in power law shape $\Gamma$, which accounted for up to 15 per cent of the variability. \\

All of the sources in our sample showed a significant third component in the long-term that is not obviously associated with a physical spectral parameter. In Section 5 we demonstrate that this arching principal component is not a physical effect at all, but rather a mathematical artefact of the PCA process caused by a lack of linearity within the data. This mathematical factor becomes much less prominent if the changes in normalization and photon index are correlated. \\

Therefore, taken at face value, the long-term variability in our sample of blazars can be described by random variations of the power law brightness and photon index, or perhaps correlated variations between the parameters with some time delay. \\

Although the results for each object were similar, the differences between the various long-term PCAs can tell us a surprising amount. In particular, the slope of the first component is stronger in objects where $\Gamma$ varies across a wider range. This slope is angled away from zero when $\Gamma$ and normalization vary independently of each other, and towards zero when they vary together. This can be used to distinguish between emission mechanisms in blazars. Blazars are known to emit in the X-ray through either synchrotron emission, the inverse Compton effect, or some combination of both according to their luminosity (Donato, 2001), and correlations between $\Gamma$ and flux are indicative of inverse Compton emission (Fatima, 2017). A decrease in the first principal component with energy can therefore be used as an indicator of Comptonization. OJ 287 is a good example of this. Its long-term PCA (Figure 3) can be explained by correlated variations in $\Gamma$ and normalization (compare to the third panel in Figure 6, and see the discussion of it in Section 5). Fatima (2017) remarks that OJ 287 is known to emit via the inverse Compton process.\\

The degree to which a power law shape dominates our results becomes obvious when compared to similar analyses of radio-quiet objects, such as many of the objects in Parker et al (2015). Radio-quiet objects display much more complicated components that can include prominent features corresponding to emission lines, absorption edges, blackbodies, and so on. While this makes PCA a useful tool for identifying model components in radio-quiet objects, it also means that it is harder to identify what each principal component corresponds to. In particular, even with just the two parameters of a power law model, a third, non-physical component is required to complete the PCA. In more complex objects with competing spectral models, it may be harder to pick out the useful results from the mathematical artefacts caused by the non-linearity of the data set. \\

The short-term PCAs are more complex and interesting when compared to the long-term results. Long-term variability is simply the sum of many smaller variations, and yet the same does not always seem to be true of the PCAs. Most observations show only a single principal component, perhaps due to limited signal-to-noise. \\

Many observations show a broken power law shape, often with an upward curve at higher energies, which is not seen in any of the long-term PCAs. This would seem to suggest a variability mechanism that only manifests over short timescales, and is washed out by larger changes in the long term, but it is unclear what this sort of mechanism could be. A future work could investigate this further using simulations with more complex models that account for factors such as shock propagations within the jet, or the influence of the AGN itself on the spectrum beyond just the jet. \\

Some observations showed no significant components at all, a sign of no rapid variability. This indicates that even highly variable objects such as blazars can show moments of steadiness, or display variability mechanisms that operate on scales greater than hours. \\

\section{Conclusions } 
PCA was used to analyse the X-ray spectra of nine blazars in order to identify variability trends across several timescales. Over long timescales, variability was found to be consistent with changes in a power law model, as should be expected in a blazar. In addition to principal components corresponding to change in normalization and $\Gamma$, a third component was seen in all objects. This component has no physical explanation, and instead was found to be a relic of the PCA process created by non-linearities within the data set. Even though each PCA shares the same broad power law shape, differences in the shapes of these components can be used to predict various qualities, such as the degree to which $\Gamma$ is varying and correlations between spectral hardness and flux. \\

Over shorter timescales, the results were more complex. Some observations contained components similar to those seen in the long-term PCAs, which over time would add up to produce the long-term variability seen in each object as one would expect. However, others showed shapes not seen in the long-term analysis, including broken power laws and a unique, curved shape with no obvious physical analogue. Most of the short-term PCAs produced only one significant component, possibly due to low signal-to-noise. The smaller number of components and less consistent results mean that it is harder to draw useful conclusions from single-observation PCAs at the moment, although there may be interesting physics to discover in this area if variability really does differ qualitatively in the short-term.\\

Principal component analysis is a useful tool that can offer a new approach with which to analyse a data set. However, even with objects as simple as blazars, it should not be trusted blindly without an understanding of its underlying assumptions and limitations. \\


\section*{Acknowledgements}

We would like to thank the reviewer for their suggestions and feedback on this work.

\appendix

\section{Specific Long-Term PCAs}
This appendix presents the results of each long-term PCA and comments on them individually. As explained in Section 3, each follows the same pattern of a normalization component, a pivoting component, and an arch-shaped non-physical component elaborated on in Section 5. These components indicate a close fit to a power law model, which is expected for blazars. Still, there are smaller differences between the objects that deserve a closer look.

\subsection{3C 273}
These principal components are the simplest and easiest to explain of any object sampled. This PCA is an excellent match to simulations of a single power law varying in both normalization and photon index, corresponding to the first and second components respectively. The third component is not physical, but rather a mathematical artefact of the PCA process. This is discussed further in Section 5. A PCA of this object appears in Parker et al (2015), but displays a slight curve in the first component. This difference is due to the difference in methods: in this work, each observation contributed only one spectrum to the long-term PCAs, whereas Parker's earlier work splits each observation into smaller sections for every PCAs. This causes their results to look like a combination of our short-term and long-term results, explaining the additional shape in the first component.
 
\subsection{H1426+428}
As with 3C 273, components corresponding to normalization and photon index can be seen clearly. Unlike 3C 273, H1426+428 shows some curvature in both major principal components, with the first increasing at high energies and the second and third suppressed at low energies. An increase in the first component can be reproduced in simulations of a power law model by increasing the degree to which $\Gamma$ varies relative to normalization, as shown in Figure 6. Larger variations in $\Gamma$ compared to normalization produce a steeper slope in the first component. A slant in the first component is seen in all of the objects, with 3C 273 being the flattest, indicating a relatively stable photon index. \\

The curved shape of the second component is harder to explain. A shape similar to this can be produced in simulations of a double power law model where two power laws vary together in normalization (Parker et al., 2015) but such a model does not reproduce the upward slope of the first component. Sambruna et al. (1997) find a variable warm absorber in the spectrum of this object (and of PKS 2155-304, which has a very similar PCA) but simulations of variable absorbers have also been unable to replicate the shapes of the first two principal components. 

\subsection{Mrk 421}
This object's PCA is unique due to the kink found shortly before 2 keV. Mrk 421 was a highly piled-up source, and it is possible that this kink is caused by instrumental features. The silicon K$\alpha$ or K$\beta$ lines could be responsible.   

\subsection{Mrk 501}
In Mrk 501, the first component again increases with energy, indicating large changes in photon index. The most notable feature is the shape of the second component, which begins rising back up around 7 keV rather than continuing downward as in the other pivoting components. The cause of this is unknown.

\subsection{OJ 287}
In OJ 287, the first principal component is decreasing with energy, rather than increasing. This can be reproduced in simulations by assuming that the variation in normalization and $\Gamma$ is correlated, rather than varying them independently of one another, as shown in Figure 6. This can explain the shape of all three significant components in this object. Enforcing a correlation between normalization and $\Gamma$ is not without precedent, and in fact seems to be the case for OJ 287 in particular. Fatima (2017) finds such a correlation, and remarks that this indicates inverse Compton emission rather than the synchrotron process. 
  
\subsection{PG 1553+113}
PG 1553+113 shows the most similarity to Mrk 501, especially in the second component. However, the first component flattens out at high energies rather than continuing to increase, and the third component is much flatter, barely showing any shape at all. 

\subsection{PKS 2155-304}
This PCA is nearly identical to that of H1426+428. Both objects are known to have warm absorbers (Sambruna 1997), but no absorption model has reproduced these principal components as of yet. 

\section{Specific Short-Term PCAs}
Here, the results of each single-observation PCA are presented individually. They are not as similar as the long-term ones, falling into the categories discussed in Section 4. 

\subsection{3C 273}
3C 273 had by far the highest number of suitable observations for this analysis, and the results are wide-ranging. Observations 012670301 and 0159960101 are shaped like a broken power law varying only in $\Gamma$. Observation 0414190101 looks like a change in normalization. Observation 0136550101 shows a pivoting and normalization component, but the pivoting component is more significant than the other. This only occurs in one other object, S50716+714. Finally, observations 0126700801 and 0651610101 show a strange, upward-curving shape not seen in any of the long-term observations. The cause of this shape is unknown, and simulations have failed to replicate it, but it also appears in short-term PCAs of 3C 279, H1426+428, and OJ 287. This unique shape could indicate the existence of a process that drives short-term variability but has no effect in the long term, but it is hard to conceive of such a process, especially for objects with as simple a spectrum as blazars.   

\subsection{3C 279}
There are not enough \xmm observations of this object to perform a long-term analysis, but one observation long enough for short-term analysis does exist. The results are similar to those for 3C 273 observations 0126700801 and 0651610101, an upward-curving slope with no obvious explanation.

\subsection{H1426+428}
H1426+428 observation 0111850201 displays the upward curving shape seen before. There also appears to be some amount of shape to the second component, but it is of very low significance and is unlikely to be a real effect. A similar shape is seen in OJ 287 observation 0761500201.

\subsection{Mrk 421}
Unlike most of the other objects, both observations of Mrk 421 show more than one principal component. Observation 0099280201 has a normalization and a pivot component, as seen in most of the long-term observations. Observation 0136540701, on the other hand, looks like a broken power law. It is worth noting that the third (nonphysical) component seen in the long-term PCAs is not discernible in observation 0099280201. This is most likely because the second component is very weak.

\subsection{OJ 287}
The upward-sloping shape is seen again in the first principal component of observation 076500201, and, like in H1428+428, the second component seems to have a slight upward shape to it near the end as well. As with H1426+428, the second component is not significant at all, but the same semblance of a shape appearing twice is unlikely to be a coincidence. It may be an artefact of the PCA process, much like the third component seen in the long-term analyses. 

\subsection{PKS 2155-304}
Observation 0124930301 indicates normalization changing alone, whereas observation 0124930601 seems to indicate changes in both normalization and $\Gamma$. 

\subsection{S50716+714}
Like 3C 279, there were not enough observations to perform a long-term PCA on S50716+714. Observation 0502271401, the only one available for this analysis, shows a pivoting component that is more significant than its normalization component. This is only seen in one other case, 3C273 observation 0136550101. It is unknown why the pivoting component would suddenly be stronger in these cases.

\end{document}